\documentclass[sigplan,10pt]{acmart}
\usepackage[T1]{fontenc}
\usepackage[utf8]{inputenc}
\usepackage{hyperref}
\usepackage{graphicx}
\usepackage{booktabs}
\usepackage{textcomp}
\usepackage{array}
\usepackage[caption=false,font=footnotesize]{subfig}
\usepackage{pifont}

\newcommand{\myitem}[1]{\smallskip\noindent\textbf{#1}}

\newcommand{\todo}[1]{}
\newcommand{\sysName}{Flowcorder}
\newcommand{\system}{{\small\textsf{\sysName}}}
\newcommand{\Cword}[1]{\mbox{\texttt{\small #1}}}

\newcommand{\MPTCP}{\textsc{mptcp}}
\newcommand{\IETF}{\textsc{ietf}}
\newcommand{\TCP}{\textsc{tcp}}

\newcommand{\DNS}{\textsc{dns}}
\newcommand{\RTT}{\textsc{rtt}}
\newcommand{\RTTs}{\textsc{rtt}{\footnotesize s}}
\newcommand{\HTTP}{\textsc{http}}
\newcommand{\IPv}[1]{\textsc{ip}{\footnotesize v#1}}
\newcommand{\IPFIX}{\textsc{ipfix}}
\newcommand{\QUIC}{\textsc{quic}}
\newcommand{\KPIs}{\textsc{kpi}{\footnotesize s}}
\newcommand{\ASes}{\textsc{as}{\footnotesize es}}
\newcommand{\eBPF}{{\footnotesize e}\textsc{bpf}}

\newcommand{\inputtikz}[1]{%
        \includegraphics{#1.pdf}
    }
\definecolor{mOrange}{HTML}{FFDCB8}
\definecolor{mGreen}{HTML}{D8FC7A}
\definecolor{mRed}{HTML}{DE2D26}
\definecolor{mGray}{gray}{0.75}
\definecolor{dGray}{gray}{0.5}
\definecolor{mBlue}{HTML}{08519C}
\definecolor{mHighlight}{HTML}{00f8dd}
\definecolor{mPurple}{HTML}{c51b8a}
\definecolor{switch}{HTML}{006996}
\colorlet{darkgreen}{black!40!green}

\renewcommand\footnotetextcopyrightpermission[1]{}

\author{Olivier Tilmans}
\authornote{O.~Tilmans is supported by a grant from F.R.S.-FNRS FRIA}
\affiliation{\institution{Université catholique de Louvain}}
\email{olivier.tilmans@uclouvain.be}

\author{Olivier Bonaventure}
\affiliation{\institution{Université catholique de Louvain}}
\email{olivier.bonaventure@uclouvain.be}

\title{COP$^2$: Continuously Observing Protocol Performance}

\begin{document}

\begin{abstract}
As enterprises move to a cloud-first approach, their network becomes crucial
to their daily operations and has to be continuously monitored. Although
passive monitoring can be convenient from a
deployment viewpoint, inferring the state of each connection can cause them to
miss important information (e.g., starvation). Furthermore, the increasing
usage of fully encrypted protocols (e.g., \QUIC{} encrypts headers), possibly over
multiple paths (e.g., \MPTCP), keeps diminishing the applicability of such
techniques to future networks.

We propose a new monitoring framework, \system, which leverages information
already maintained by the end-hosts and records Key Performance Indicators
(\KPIs) from their transport protocols. More specifically, we present a generic
approach which inserts lightweight \eBPF{} probes at runtime in the protocol
implementations. These probes extract \KPIs{} from the per-connection states,
and eventually export them over \IPFIX{} for analysis.

We present an application of this technique to the Linux kernel \TCP{} stack
and demonstrate its generality by extending it to support \MPTCP{}. Our
performance evaluation confirms that its overhead is negligible. Finally, we
present live measurements collected with \system{} in a campus network,
highlighting some insights provided by our framework.
\end{abstract}

\maketitle

\section{Introduction}

Network performance depends on a variety of factors such as link delays
and bandwidth, router buffers, routing or transport protocols.
Some of these are controlled by the network operators, others by
the end-hosts. To detect potential issues, and ensure their proper operations,
most network operators monitor a wide range of statistics on the health of
their networks,
which can be classified in three categories.
First, health metrics capture the status of network elements. Most networks
record those using \textsc{snmp}, polling their devices every few
minutes to collect various statistics (e.g., link load,
\textsc{cpu} usage, size of forwarding tables).
Operators often also collect statistics about the traffic itself,
usually using NetFlow/\IPFIX{}~\cite{santos2015network,
hofstede2014flow,trammell2011introduction}. These provide more detailed
 information about the flows crossing the network (e.g., layer-4 5-tuples,
volumes in bytes and packet), and enable various management
applications~\cite{li2013survey} (e.g., identifying major source/destination
pairs~\cite{yeganeh2017view}, heavy-hitters~\cite{gangam2013pegasus}, or
detecting DDoS attacks~\cite{sekar2006lads,van2015real}).
Finally, operators monitor key performance metrics which are important for many
end-to-end applications, such as delays, packet losses, and retransmissions.
On one hand, active measurements techniques~\cite{ipsla,luo2009design} collect
these metrics by generating test traffic (e.g., pings). On the other hand,
passive measurements~\cite{finamore2011experiences,john2010passive} infer these
performance metrics by analyzing the packets that traverse the
network (e.g., using network taps which maintain per-flow states to accurately
measure Round-Trip-Times (\RTT), retransmissions, packet losses and
duplications~\cite{Mellia_tstat:2002}).


Although widely deployed, passive monitoring suffers
from several important limitations. First, as link speeds
increase, it becomes more and more difficult to maintain the per-flow
state that is required to collect detailed performance
metrics~\cite{Trevisan_Traffic:2017}. Second, as multipath protocol deployment
increases (e.g., \MPTCP{}~\cite{RFC6824} is used in iPhones~\cite{ios-mptcp}
and for other services~\cite{bonaventure2016multipath}),
passive monitors only see
a subset of the packets belonging to a connection. This compromises their
ability to operate properly~\cite{blackhat-mptcp}. Finally, the most important
 threat against the passive collection of network performance metrics is the
deployment of encrypted protocols, such as
\QUIC{}~\cite{langley2017quic}. \QUIC{} replaces the \textsc{htt/tls/tcp} stack with a simpler protocol that runs over
\textsc{udp}. Google
estimates~\cite{langley2017quic} that \QUIC{} already represents more than 7\%
of the total Internet traffic. Recent measurements indicate that content
providers have started to deploy \QUIC{} massively~\cite{ruth2018first}.
The \IETF{} is currently finalizing a standardized version of
\QUIC{}~\cite{quic-wg}.

From a performance monitoring viewpoint, an important feature of \QUIC{}
is that all the payload and most of the header of the packets
are encrypted. This prevents the middlebox ossification problems that
affect protocols such as \TCP~\cite{honda2011still,papastergiou2017ossifying},
but it also greatly decreases the ability for network operators to monitor
network performance. This prompted some of them to ask to modify \QUIC{} to be
able to extract performance information from its
headers~\cite{Stephan_QUIC:2017}. The \IETF{} answered those
operational concerns by reserving one bit in the \QUIC{} header (the spin-bit~\cite{Trammel_spin:2018}), exposing
limited delay information. Multipath extensions
to \QUIC{} have already been proposed~\cite{viernickel2018multipath,de2017multipath}.

To keep collecting end-to-end performance metrics of their users flows,
enterprise network operators need a different approach than passive monitoring
to be future proof.



\myitem{Problem statement}
\emph{How can we support the legitimate need of fine grained
performance information from enterprise network operators in presence of
encrypted, multipath protocols?}

\begin{figure}
    \subfloat[Most networks are monitored using passive measurements made by
    the network devices.\label{fig:monitor_base}]{%
        \inputtikz{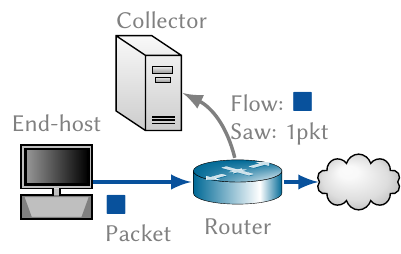}
    }\hfill\subfloat[\sysName{} exports information from the end-hosts about
        their connection performance.\label{fig:monitor_system}]{%
        \inputtikz{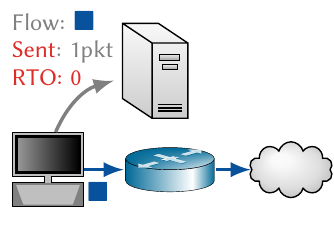}
    }
    \caption{\sysName{} enhances network monitoring with fine-grained
    measurements about connections.}\label{fig:architecture}
\end{figure}

\myitem{Key challenges} Designing a monitoring framework that answers this
question raises at least four challenges.
First, this framework must accurately depict the performance experienced by
the end-hosts. This limits the applicability of active measurements, as this
might hide issues specific to the used protocol (e.g., \TCP{} \textsc{rto}).
Second, it must support multipath protocols, and thus monitor the
performance of all paths used by a given connection. This limits the
possibility of using passive monitoring since this would require
coordination among the monitors located on
different paths. Third, supporting encrypted protocols
prohibits such framework from analysing packet headers or contents and
prohibits the utilisation of ``transparent'' proxies.
Finally, it should operate with a low overhead, limiting the generated
statistics to the minimum to establish a baseline for normal operation, while
also enabling to quickly capture and detect performance issues.

\myitem{\sysName{}} We introduce \system, a novel enterprise network monitoring
 framework which addresses the above challenges. The key insight behind
 \system{} is to leverage the per-connection information that is already
 maintained by the end-hosts themselves.

Instrumenting the transport stacks of the end-hosts enables \system{} to
compute Key performance Indicators (\KPIs) for each connection. By capturing
such \KPIs{} at specific moments of the connection life-cycle, \system{} can
 then build \emph{performance profiles} of connections. Finally, \system{}
aggregates those profiles and exports them over \IPFIX{}, integrating with
 existing monitoring infrastructure and enabling analyzes across hosts,
protocols, remote services and/or \textsc{isp}s.

\myitem{Contributions} Our main contributions are:
\begin{itemize}
    \item A novel enterprise monitoring framework to monitor the network performance
        experienced by the end-hosts~(\S\ref{sec:overview}).
    \item A generic approach to export performance profiles of connections
        by transparently extracting \KPIs{} from existing protocol
        implementations~(\S\ref{sec:perf_profile}).
    \item An application of the approach to realize an event-based
        instrumentation of the Linux kernel \TCP{} stack~(\S\ref{sec:tcp}),
        a demonstration of the generality of the approach by extending it to
        support \MPTCP~(\S\ref{subsec:mptcp}), and an evaluation showing
        its low-overhead~(\S\ref{sec:eval}).
    \item A case study highlighting insights provided by \system{} when deployed
        in a campus network~(\S\ref{sec:case_study}).
\end{itemize}

\begin{figure*}[t]
    \inputtikz{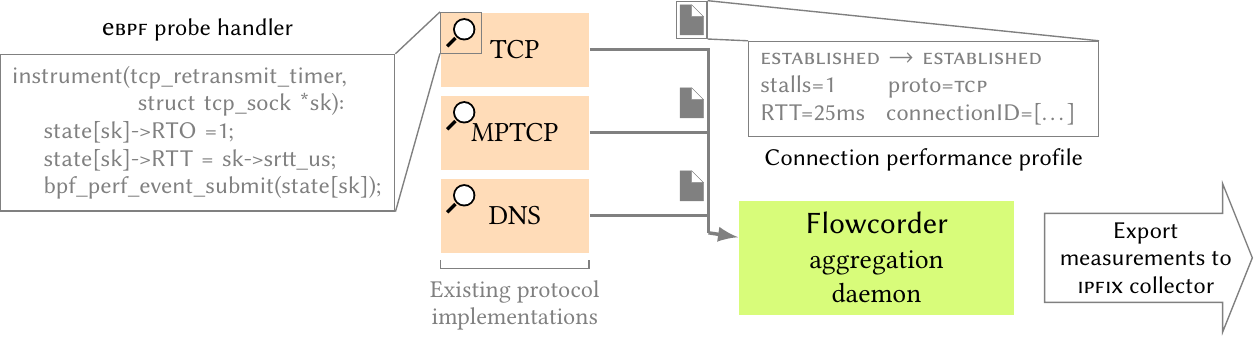}
    \caption{\sysName{} enables to evaluate network performance from generic
        Key Performance Indicators collected on the end-hosts for every
    connection.}\label{fig:overview}
\end{figure*}
\section{\sysName}\label{sec:overview}

Many networks monitor their traffic using in-network appliances that inspect
packets crossing them, and eventually export statistics to measurement
collectors using a protocol such as \IPFIX{} (Fig.~\ref{fig:monitor_base}).
While sufficient to track traffic demands, or collect rough traffic statistics
through passive inference of the connection states, these techniques hardly
scale if the operators requires fine-grained performance measurements on a
per-connection basis. \system{} instead pushes the monitoring processes directly
on the end-hosts (Fig.~\ref{fig:monitor_system}). By monitoring the
per-connection states, \system{} can then record the performance of the
connections, as experienced by the end-users, and then export those over
\IPFIX{} to complement existing measurement infrastructure. The rest of this
section illustrates the different building blocks
making up \system{}, visible on Fig.~\ref{fig:overview}. More specifically, we
consider a network administrator who wants to use \system{} to answer
the following a high-level question: \emph{``Which provider performs the best to
connect to a remote storage service accessed over \TCP{}?''}

\myitem{Computing performance profiles.}
The first step to answer this high-level question is to identify
\KPIs{}~(\S\ref{subsec:what_perf}) that enable to characterize the
performance of the instrumented protocol. Such \KPIs{} should contain
general statistics about the connection, as well as metrics indicating possible
performance issues, specific to the protocol.

For example, high-level \KPIs{} to answer our illustrative question could be:
$(i)$ the number of bytes transferred and assumed to be lost;
$(ii)$ the amount of reordering~\cite{jaiswal2007measurement,augustin2011measuring,bellardo2002measuring}
that occurred in the network;
and $(iii)$ signs of bufferbloat, such as the number bytes received multiple
times, thus signaling a retransmission timeout on the sender, or times where
the connection stalled and was blocked from sending pending data for several
\RTT{}s (\TCP{} \textsc{rto}).

Continuously streaming the collected \KPIs{} is inefficient as, beside wasting
resources, it might hide the key performance outliers in the noise generated
by the huge number of smaller variations. Instead, \system{} exports the
\KPIs{} of a connection only at specific moments in the connection
life-cycle~(\S\ref{subsec:fsm_perf}). In-between these exports, the \KPIs{} are
buffered in a lightweight aggregation daemon, local to the end-host. Once the decision
to export the measurement is made, this aggregation daemon computes a
\emph{performance profile} of the connection: statistics computed over \KPIs{}
(e.g., moving averages, counter increase) during well-defined moments of the
connection life-cycle. The performance profile is then serialized as an
\IPFIX{} record and added in a pending \IPFIX{} message buffer. As we want to
minimize the processing load on the collector and take advantage of the
features provided by \IPFIX{}, the message is only exported once its size
reaches the local \textsc{mtu}.

In our example, a connection towards the remote storage service that would
experience one retransmission timeout in its entire life-cycle would generate
four performance profiles: $(i)$ one describing the connection establishment;
$(ii)$ one describing the performance of the data transfer (e.g., average \RTT{},
byte counters, number of \textsc{rto} experienced) up to the \textsc{rto};
$(iii)$ one describing the performance while the connection is considered as
lossy; and $(iv)$ a final one describing the performance since the end of the
lossy state and how the connection ended (e.g., did it abruptly end with a
\TCP{} \textsc{rst}?).

\myitem{Collecting \KPIs.}
Under the hood, \system{} instruments existing transport protocol implementations
on the end-hosts. Many methods exist to collect such statistics, such
as extracting them from a general purpose
loggers~\cite{windows-ewt,chrome-logger} or polling~\cite{snmp}.
Instead, \system{} uses an event-based method. More
specifically, \system{} inserts \eBPF{} probes at specific code paths in the
transport protocol implementations~(\S\ref{sec:ebpf}).
When the end-host stack reaches one of
these probes, the probe handler is executed, computes \KPIs{} of the
connection, exports them in an asynchronous channel to the aggregation daemon,
and then resumes the normal execution of the protocol implementation.
Beside minimizing the instrumentation overhead~(\S\ref{sec:eval}), this
approach is also extremely flexible as it does not require any
support from the implementation (e.g., \textsc{mib}s), and is thus not
restricted to a predefined set of metrics, computed in an opaque manner.

In the example of Fig.~\ref{fig:overview}, we see that one such probe has been
setup to intercept the expiration of the \TCP{} retransmission timer. If any
connection experiences a \textsc{rto}, this handler then increases the
\textsc{kpi} counting \textsc{rto}'s and updates the connection's \RTT{}
estimated by \TCP{}, then exports it for processing in user-space.

\myitem{Analyzing performance profiles.}
\system{} produces measurements that can be collected, parsed and analyzed by
any IPFIX collector supporting custom Information
Elements~\cite{ipfix-custom-ie}.
Performance profiles are independent views of the performance of a connection
during a given window of time, and one can be analyzed separately from the
others belonging to the same connection. These performance profiles thus enable
the network operator to build several views of the network according to key
metrics using simple database queries, and to analyze them~(\S\ref{sec:case_study}).

For example, to answer his question, our network administrator could compute
generic statistics such as mean, variance and median of all performance profiles
contained in a given time window, aggregated by provider, and run
hypothesis tests. These results could also be split based on the IP version,
or compared against the general trend to access all other remote services.
Finally, beside numerical tests, one can also generate time series and plot
them in monitoring dashboards.

\section{Recording protocol performance}\label{sec:perf_profile}

\system{} records\emph{performance profiles} of connections
directly on end-hosts, and
exports them to a collector for further analysis.
Achieving this requires addressing three issues:
$(i)$ What should a performance profile contain to describe a connection and
indicate performance issues (\S\ref{subsec:what_perf})?;
$(ii)$ How can we collect these key metrics from the protocol implementations?;
and $(iii)$ When should these profiles be computed to maximize the
accuracy of the
measurements while minimizing the overhead of \system{} (\S\ref{subsec:fsm_perf})?

\subsection{Characterizing protocol performance}\label{subsec:what_perf}

\def\figWidth{\columnwidth}
\begin{table}
    \footnotesize\centering{}
    \renewcommand{\arraystretch}{1.2}
    \caption{Key Performance Indicators can answer most questions about
 transport protocol performance}\label{table:kpi}
    \begin{minipage}{\figWidth}
        \renewcommand{\thempfootnote}{\fnsymbol{mpfootnote}}
        \begin{tabular}{@{}l@{}lp{.7\figWidth}@{}}\toprule
            & \textbf{KPI} & \textbf{\small{}Description}\\
            \midrule
            $\sum\footnote{$\sum$ denotes a counter over a time window}$
            & \Cword{Sent} & Data\footnote{Most \KPIs{} can be duplicated to track
                byte-counts and packets ('data')} sent towards the remote host\\
            $\sum$ & \Cword{Received} & Data received and processed by the end host\\
            $\sum$ & \Cword{Lost} & Data assumed to
                be lost in the network\\
            $\sum$ & \Cword{Errors} & Data received corrupted\\
            $\mathcal{A}$\footnote{$\mathcal{A}$ denotes an average and a variance over a time window}
            & \Cword{RTT} & Mean Round-Trip-Time and variance (i.e., jitter)\\
            $\sum$ & \Cword{Duplicates} & Received data already acknowledged\\
            $\sum$ & \Cword{OFO} & Data received out-of-order\\
            $\mathcal{A}$ & \Cword{OFO-dist} & Distance of out-of-order data from the expected one\\
            $\sum$ & \Cword{Stalls} & Count when the connection delays the sending of any pending
            data during several \RTTs{}\\
            \bottomrule
        \end{tabular}
    \end{minipage}
\end{table}

Connection-oriented transport protocols such as \TCP{} maintain
state and usually expose some debugging information (
e.g. \Cword{struct tcp\_info}~\cite{tcpinfo} on Linux or macOS).
However, recording the entire state for each established connection is
impractical. Most of this information is very specific to the protocol
implementation and does not always relate to connection performance.
For example, one can find the distance (in terms of \TCP{} segments) between
the last out-of-order segment and the expected sequence number or the value of
the slow-start threshold in the \Cword{struct tcp\_info}, both of which give
almost no insight to qualify the connection performance. Finally, while
\system{} aims to collect fine-grained measurements about protocol performance
as experienced by the end-hosts, recording every single data point would be
counter-productive, as the more critical observations will end up buried in a
huge pile of data.

Instead, we characterize protocol performance by recording the evolution
of Key Performance Indicators (\KPIs) during a connection. Example \KPIs{} are
listed in Table~\ref{table:kpi}. Recording \Cword{Sent} and \Cword{Received}
bytes quantifies the volume transported on a connection, while
tracking the number of segments quantifies the packet rate (e.g., an interactive
\textsc{ssh} session produces many small \TCP{} segments). Recording
\Cword{Lost} segments or segments with a checksum error
(\Cword{Errors}), enables to qualify the path used by the connection. Tracking
the evolution of the \RTT{} (and thus implicitly its jitter) can be
used to estimate whether congestion is building up in the network (and is the main
source of information of some congestion control algorithms such as
\textsc{bbr}~\cite{tcp-bbr}). Similarly, recording the reception of segments
containing already acknowledged data is an indication that the remote host
mistakenly assumed their loss, which could be a sign of a possible bufferbloat. Measuring
the amount of packet reordering is also useful, especially in the context of
transport protocols, as its occurrence often limits the maximum achievable
throughput. Finally, recording when a connection is prevented from making
progress is a strong signal that something bad happened in the network (e.g.,
triggering a \TCP{} \textsc{rto}).

From these \KPIs{}, network administrators can then answer complex high-level
questions characterising the performance of the network, such as: $(i)$ what
is the best response time that can be expected when connecting to a remote server?;
$(ii)$ Is the connection suitable for bulk transfers?; or $(iii)$ Is
the network congested?

\subsection{Collecting KPIs{} from implementations}\label{sec:ebpf}

Recording the evolution of the \KPIs{} of a connection on the end-hosts requires
to extract them directly from the protocol implementation. Achieving this is usually
possible using poll-based techniques. For example, \textsc{snmp} can be
used to query the \TCP{} Management Information
Base~(\textsc{mib})~\cite{tcp-mib}. Some OS'es also define APIs to retrieve
information~\cite{tcpinfo,tcpmetrics}, or log events to a centralized
journal~\cite{windows-ewt} which can then be monitored.

These techniques however come with two limitations.
First, the information they give is limited to the explicitly defined metrics.
For example, counting \TCP{} out-of-order packets, as well as characterizing their
out-of-order distance is impossible on Linux with the existing \textsc{api}. Counting
received duplicates is not feasible either. Second, by requiring the monitoring
tool to poll them, getting more accurate information about performance changes
imposes a polling frequency and thus a high resource usage on the end-hosts. For example, characterizing the connection
establishment times requires to precisely track the first few packets of a
connection, which could be exchanged within a few milliseconds.

To address these issues, \system{} bypasses these traditional techniques, and
directly instruments the protocol implementation at runtime.

\myitem{Dynamic tracing using \eBPF.}
\system{} leverages the existing dynamic tracing tools such as
\emph{kernel probes}~\cite{kprobe-lwn}, or DTrace~\cite{dtrace}.
These enable
to insert lightweight probes at runtime at arbitrary locations in
either kernel~(e.g.,
to instrument the \TCP{} kernel implementation~\S\ref{sec:tcp}) or
user-space code~(e.g., to instrument \DNS{} resolution routines, for which we
present collected measurements in~\S\ref{sec:case_study}), typically around function
calls. Conceptually similar to breakpoints and debugging watches, these probes
automatically call user-defined handlers before and after executing
the probed
instruction. These handlers have complete access to the memory,
as well as to the content of the \textsc{cpu} registers. More
recently, the Linux kernel added code to
define such handlers using extended Berkeley Packet Filters
(\eBPF{})~\cite{eBPF}.

\eBPF{} code is pre-loaded in the kernel using the \Cword{bpf()}
system call. This \eBPF{} code is executed in an in-kernel virtual
machine that mimics a RISC 64-bits \textsc{cpu} architecture, with 11
registers and a 512 bytes stack. This code can be interpreted, but
many architectures include a \textsc{jit} that compiles the \eBPF{} bytecode.
Before accepting to load an \eBPF{} code, a verifier ensures safety guarantees
such as proof of termination (e.g., by limiting the overall number of
instructions and disallowing non-unrollable loops) and checks memory-access.
\eBPF{} code executed within the kernel can asynchronously communicate with
user-space processes using perf events (FIFO queues). Additionally, \eBPF{}
programs can defines maps, which let them maintain state in-between executions.
When an \eBPF{} probe handler is executed, it receives an instance of the
\Cword{struct pt\_regs}, which describes the content of the \textsc{cpu}
registers when the probe was hit, including the value of the
stack pointer. This enables the \eBPF{} handler to inspect the
function arguments, or to explore
the memory of the instrumented code.
These capabilities make eBPF a target of choice to write probe handlers, as
they guarantee that the handlers will not cause crashes nor hang the
instrumented code, while also enabling it to compute complex statistics and
easily report them to user-space.

This approach has at least five advantages.
First, by leveraging state transitions that are internal to the implementation,
it ensures an accurate translation to \KPIs. For example, by recording
retransmission timer expirations, it easily distinguishes between
a connection that had no data to send for a while and a connection that was
stalled and had to wait a complete RTO before sending anything else.
Second, it seamlessly adapts to settings local to the host --- for example,
the \TCP{} duplicate \textsc{ack} threshold, or the support for SACK on a per connection
basis --- that alter the behavior of the transport protocol. As such, it
accurately captures the performance experienced by all instrumented end-hosts.
Third, as it implements a push-based model where the transport stack itself
calls \system{}, it minimizes the overhead on the end-hosts. Indeed, as the
probe locations guarantee that all \textsc{kpi} changes will be detected, this
avoids the need for constant, high-frequency, polling of the state-variables.
Fourth, as it enables to both read per-connection states and to compute
arbitrary statistics that can be stored in maps (thus defining custom ancillary
state), this approach is highly flexible, as it does not rely on
specific support from the protocol implementation.
Finally, it could also be applied to encrypted transport protocols such as QUIC
since it does not use the packet data but instead the state-variables of the
protocol implementation.

\subsection{Creating performance profiles}\label{subsec:fsm_perf}

\begin{figure}
    \centering{}
    \inputtikz{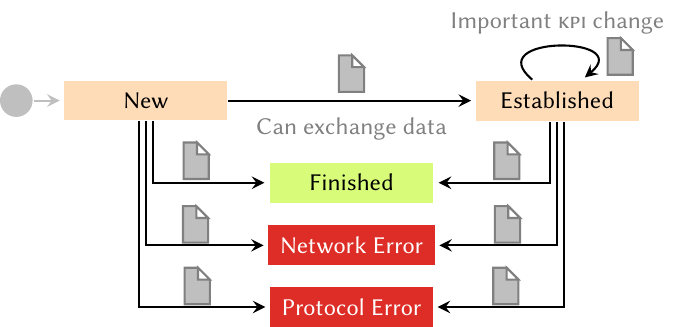}
    \caption{Transport protocol flows can be abstracted in a general FSM, where
        state-transitions act as signal to compute performance
	profiles.}\label{fig:fsm}
\end{figure}

To use dynamic tracing and \eBPF{} handlers to instrument a particular transport
protocol, one needs to pick probe insertion locations to catch updates to the
state of a connection.
While a straw-man approach would pick the main functions involved
in every send and receive operation, and continuously stream the connection
\KPIs{} after each sent and received packet, this would impose a high overhead
without necessarily providing useful measurements. Indeed, once the probes are
inserted, their handlers are executed for every connection hitting that code
path. Instead, we aim at recording the evolution of \KPIs{} between key events
in the connection life-cycle. To this end, we place probes at locations that
are seldom reached, yet catch all important events affecting the connection,
and record statistics describing the evolution of the \KPIs{} between two
events. We call such set of statistics the \emph{performance profile} of a
connection.

A first set of events are defined by the protocol specifications. Such
specification is usually composed of two different parts. The first
is the syntax of the protocol messages, which can be expressed
informally with packet descriptions or more formally by using a grammar (e.g.,
\textsc{abnf}~\cite{rfc5234}, \textsc{asn}.1~\cite{asn1}). The second part of
the specification describes how and when these messages are sent and processed.
Most Internet protocols specifications use Finite State Machines (\textsc{fsm})
to represent the interactions among the communicating hosts.
Although implementations are usually
not directly derived from their specification (e.g. for performance
reasons or
ease of maintenance), most implementations also include the key states and
transitions of the protocol specifications. For example, most \TCP{}
implementations include the \Cword{SYN\_RCVD,
SYN\_SENT} and \Cword{ESTABLISHED} state of the TCP
specifications~\cite{rfc793}. While state transitions signal that a connection is
making progress, not all of them provide similar information (e.g., transitions
into the \TCP{} \Cword{TIMEWAIT} state give no information on the connection
besides that ''it is about to close``). Ultimately, these \textsc{fsm} describe the
life-cycle of a connection. They can thus be abstracted by mapping their state
and transitions to the three key phases in a connection life-cycle:
$(i)$ the connection establishment; $(ii)$ the exchange of data; and
$(iii)$ the connection tear-down. These three stages enable us to define the
abstract \textsc{fsm} visible on Fig.~\ref{fig:fsm}. When the state of a
connection in this simplified FSM changes, it is a signal that \system{} needs
to create a \emph{performance profile} for the connection. Performance profiles
should thus also contain the start and end states corresponding to their
transition, enabling to compare the performance of connections for similar
transitions (e.g., characterize the connection establishment delay).

A second set of events that requires \system{} to generate a performance profile
are the functions in the protocol implementation that indicate that an
unexpected event occurred (e.g., a retransmission timeout). We model this
by a looping transition in the \Cword{ESTABLISHED} state in Fig.~\ref{fig:fsm}.

Finally, a third set of probe locations is defined by \KPIs{} that are not
computed by default by the protocol implementation. For example, metrics
related to reordering for the \TCP{} instrumentation. Tracking these \KPIs{}
then implies to create an ancillary state for the connection (e.g., using an
\eBPF{} map), and updating it as the connection advances.

Once exported by the \eBPF{} handlers, these performance profiles will
eventually be received by an user-space aggregation daemon. This daemon then
serialises these profiles to an \IPFIX{} record, adding in the process
information to identify both the connection (e.g., the \TCP{} 5-tuple) as well
as the network path used (e.g., the egress interface and source address). This
record is then eventually exported to the collector.

\section{Instrumenting TCP with eBPF}\label{sec:tcp}

To demonstrate the applicability of our approach, we have applied it to
the \TCP{} implementation of the Linux kernel. This is a high-performance and
widely used \TCP{} implementation that has been tuned over more than a decade.
We first introduce the \KPIs{} building up the performance profiles
of \TCP{} connections~(\S\ref{subsec:tcp-kpi}). Then, we describe the various
\eBPF{} handlers that are used, and illustrate their
interactions~(\S\ref{subsec:tcp-probes}). Finally, we present how we
have extended this instrumentation to support \MPTCP{}~(\S\ref{subsec:mptcp}),
showing the genericity and the flexibility of our approach.

\subsection{Selecting \KPIs}\label{subsec:tcp-kpi}

Instrumenting the Linux kernel \TCP{} stack requires to map the chosen
\KPIs{} to \TCP{} state variables. A \TCP{} connection is represented in the
kernel using the \Cword{struct tcp\_sock}. As-is, this structure already
contains most of the \KPIs{} presented in Table~\ref{table:kpi}. For example,
\Cword{bytes\_received} tracks the received bytes; \Cword{srtt\_us} is a
moving average of the estimated \TCP{} \RTT{}. Computing
the statistics to create a performance profile from these state variables thus
requires the \eBPF{} handler to: $(i)$ retrieve the address of the connection
state from the parameters of the instrumented functions; $(ii)$ copy the relevant state
variables from the kernel memory to the \eBPF{} stack; and $(iii)$ compute the
statistics on the evolution of the \KPIs{} that these variables represent.

Unfortunately, not all \KPIs{} from Table~\ref{table:kpi} are directly
available in the
\TCP{} implementation. More specifically, four \KPIs{} are missing. First,
the number of duplicate incast bytes (\Cword{Duplicates}) is never recorded.
If a connection receives a segment already (partially) acknowledged, the
implementation ignores its payload. Second, the number of retransmission
timeouts~(\Cword{Stalls}) is not recorded. Similarly, the number of
bytes and packets
that arrived out of order (\Cword{OFO}) is not tracked. Finally, the existing
\Cword{reordering} connection state variable is not sufficient to represent the
distance between out-of-order packets (\Cword{OFO-dist}). Indeed, while it does
express an out-of-order distance, it does so in terms of number of MSS-sized
segments, and represents only the value computed for the last packet.
Furthermore, it is clamped by a \texttt{sysctl} value.

Recording such ''custom`` \KPIs{} thus requires to create an \eBPF{} map alongside the
probe handlers. This map can then be used to contain the ancillary state for
each monitored connection (i.e., map a connection state to a data structure
containing the value of the \KPIs{} not provided by the protocol implementation).
Managing this map has two implications. First, new entries must be added for
any connection that will be monitored. This is especially important for
connections initiated by the end-host itself. Indeed, if the \textsc{tcp syn}
they send is lost, the retransmission timer will expire, and the
count of connection stalls will need to be increased. This does not apply for
inbound connection requests, as creating state before their acceptance by
user-space application would provide a Denial-of-Service attack vector.
Similarly, this ancillary state must be purged when the connection is over. The
second implication of managing such ancillary state is that it imposes to
insert \eBPF{} code at every location where one of its value needs to be updated.
Fortunately, as the missing \KPIs{} represent very specific behaviors, these
only require to instrument two extra locations
(see~\S\ref{subsec:tcp-probes}).

\subsection{Defining \eBPF{} probes}\label{subsec:tcp-probes}
\newcommand{\checked}{\rlap{$\square$}{\raisebox{2pt}{\large\hspace{1pt}\ding{51}}}%
\hspace{-2.5pt}}
\begin{table*}
    \centering{}
    \renewcommand{\arraystretch}{1.2}
    \begin{tabular}{@{}lc@{~}cp{.7\linewidth}@{}}
        \toprule
        \textbf{Probe location} & \textbf{Pre} & \textbf{Post} & \textbf{Handler description}\\
        \midrule
        \Cword{tcp\_v\emph{[46]}\_connect} & \checked{} & \checked{}
                                           & Register a new connection attempt and initialize its ancillary state; export \KPIs{} to an error state if the function returns an error which indicates
        a cancellation of the connection.\\
        \Cword{tcp\_finish\_connect} & \checked{} & $\square$ & Exports \KPIs{} indicating the
        establishment of a new outbound connection.\\
        \Cword{inet\_csk\_accept} & $\square$ & \checked{} & Exports \KPIs{} for a new inbound connection accepted by user-space.\\
        \Cword{tcp\_set\_state} & \checked{} & $\square$ & If a connection moves to \Cword{TCP\_CLOSE}, compute
        its final state and exports its \KPIs.\\
        \Cword{tcp\_retransmit\_timer} & \checked{} & $\square$ &
        Export \KPIs{} if the connection has stalled and enters a lossy state once established.\\
        \Cword{tcp\_fastretrans\_alert} & $\square$ & \checked{} & If the connection congestion control state
        moves back to \Cword{TCP\_CA\_OPEN} (e.g., has recovered from an \textsc{rto}), exports
        \KPIs{} to mark the end of the lossy state.\\
        \Cword{tcp\_validate\_incoming} & \checked{} & \checked{} & Detect incast duplicates; update
        the reordering \KPIs{} if the packet enters the \Cword{ofo\_queue}.\\
        \bottomrule
    \end{tabular}
    \caption{A few probes in the Linux TCP implementation act as events to detect many performance changes.}\label{table:tcp_functions}
\end{table*}

Table~\ref{table:tcp_functions} lists the functions of the Linux kernel
where we insert our probes as well as their handler(s).
These functions were chosen to minimize the overhead induced by the probes,
i.e., they are never executed in the context of the \TCP{}
``fast-path'' processing. They fall into two categories. First,
we instrument the functions that correspond to state changes in the \TCP{} \textsc{fsm}
(i.e., from \Cword{tcp\_v6\_connect} to \Cword{tcp\_set\_state}).
These indicate changes in the connection life-cycle and thus mandate to compute \KPIs{}.
Second, we instrument functions that denote events which require us to update
our ancillary connection state.
 More specifically,
\Cword{tcp\_retransmit\_timer} let us track expirations of the retransmission timer.
If a connection experiences a RTO, and its write queue is not empty or the
user-space is blocked on a syscall, then it means that the connection has stalled.
\Cword{tcp\_fast\_retrans\_alert} may signal that a connection has
recovered from a RTO (i.e., that the network is stable again) and moved back in
the established state.
\Cword{tcp\_validate\_incoming}'s instrumentation is split into two handlers.
First, it detects whether an incoming segment has already (partially) been
acknowledged. Such a segment is an explicit signal that the other host
experienced a retransmission timeout.
Second, if the function accepts the received segment, this means that it is an
out-of-order segment, and the handler updates the statistics tracking the reordering.
Furthermore, as both \Cword{tcp\_retransmit\_timer} and
\Cword{tcp\_fast\_retrans\_alert} indicate that a significant performance event
has occurred (a succession of losses in the network, and then a recovery),
their handler also export \KPIs{}. This eventually creates performance profiles
looping on the \Cword{ESTABLISHED} state, enabling to describe the performance
of the connection before, during, and after such transient events (e.g., a
flash crowd causing congestion).

\begin{figure}
    \centering{}
    \inputtikz{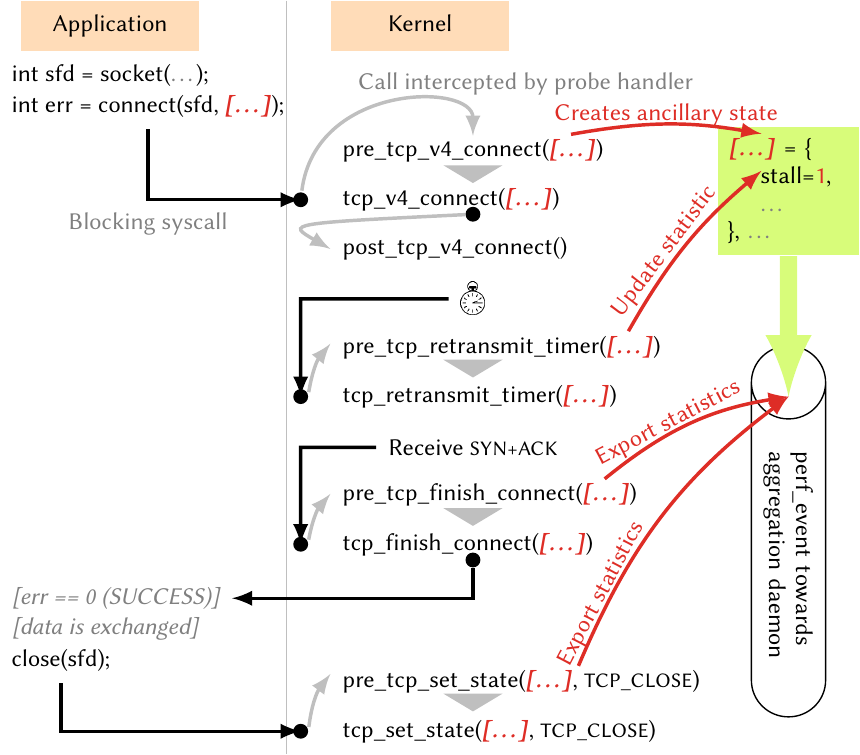}
    \caption{Abstract time-sequence diagram of the generated performance
        profiles of a TCP connection which loses its initial SYN, exchanges
        data, then closes. With a few kernel probes, our
        eBPF handlers trace the entire connection life-cycle and report it to an
        user-space daemon.}\label{fig:time-seq-connect}
\end{figure}

\myitem{Collecting \KPIs{} for a new outbound connection.}
We now illustrate how \system{} exports \KPIs{} describing the
establishment of a new outbound
connection.
In the example shown in Fig.~\ref{fig:time-seq-connect}, an
application creates a regular \TCP{} socket. Then,
it tries to establish a \TCP{} connection with the
\Cword{connect()} system call. This system call is processed by
the kernel, and eventually reaches the \Cword{tcp\_v4\_connect()}
function, for which \system{} had registered a probe. This
probe is executed before the instrumented function. It registers basic
information about this connection establishment, such as its destination address and
the time at which it started. Then, the kernel executes the
\Cword{tcp\_connect()} function, eventually sending a \TCP{} \textsc{syn} segment.
When the function exits, the post handler is executed and immediately returns
as the connection was successfully initiated and the kernel switches to other
tasks. Unfortunately, this initial \textsc{syn} does not reach the destination.
After some time, the retransmission timer expires. This causes the
kernel to execute the \Cword{tcp\_retransmit\_timer()} function. Again,
\system{} intercepts that call using a probe, which increments the number of
stalls. The kernel then sends a second \TCP{} \textsc{syn}.

When receiving
the corresponding \textsc{syn+ack}, the kernel reaches
\Cword{tcp\_finish\_connect()}. As its corresponding \eBPF{}
handler is awoken, \system{} marks the connection as established, computes its
\KPIs{} and sends them to the user-space aggregation daemon using a
\Cword{perf\_event}. This daemon asynchronously fetches and analyzes the
\KPIs{}, builds the performance profile of this new connection and adds it in
its \IPFIX{} pending message buffer to send it later to the collector.
In parallel, the \Cword{tcp\_finish\_connect()} kernel function completes
and wakes up the application which can use the connection.

If the network then behaves perfectly (e.g., no reordering, and no losses), the
probes placed in the kernel are never reached thus never
executed for that connection. Finally, when the application closes its socket,
the kernel eventually calls \Cword{tcp\_set\_state} to move the underlying
connection to the \Cword{TCP\_CLOSE} state. \system{} intercepts this call,
computes the final set of \KPIs{} for this connection, and exports a performance
profile covering the entire connection and reaching a final state describing how
the connection ended (e.g., \textsc{finished} if both \TCP{} \textsc{fin}'s
were received and acknowledged).

\subsection{Supporting \MPTCP}\label{subsec:mptcp}

\MPTCP{} is a new \TCP{} extension which enables to
operate a single \TCP{} connection over multiple paths \cite{RFC6824}. Two main
implementations of this protocol exists: the reference one in the Linux
kernel~\cite{multipath-tcp-org} and one deployed by Apple on
iOS~\cite{ios-mptcp}. We now demonstrate the genericity of \system{}, by
enabling it to record performance profiles of \MPTCP{} connections.

To instrument \MPTCP{}, a few architectural details have to be taken into
account. Despite being a relatively complex implementation
(${\sim}18$kLOC), it is heavily tied to the existing \TCP{} implementation.
At its heart, a \MPTCP{} connection operating over two paths is composed in the
kernel of two \TCP{} connections, and of one meta-socket. This meta
socket is the one exposed to user-space. It hijacks the socket \textsc{api}
used by \TCP{} (i.e., user-space programs use \MPTCP{} by default).
Sending data using \MPTCP{} requires to break the bytestream received on the
meta-socket into chunks with a \MPTCP{} sequence number (\textsc{dss}), and then
to send those over one of the subflows. The receiver's meta socket then reads
the receive queues of its subflows, and reassembles the original bytestream thanks
to the \textsc{dss}.

Instrumenting this implementation poses three challenges: $(i)$ differentiating
between a new \MPTCP{} connection and regular \TCP{} one can only be done once the
\textsc{syn+ack} has been received, since \MPTCP{} connection will contain a
dedicated option (\textsc{mp\_capable}); $(ii)$ \MPTCP{} subflows will trigger
the same \eBPF{} probes as regular \TCP{} connections; $(iii)$ new subflows
can be created directly by the meta-socket.

\myitem{\KPIs{} specific to \MPTCP{}.} As \MPTCP{} subflows operate as regular
\TCP{} connections, we use the same set of \KPIs{} as in \S\ref{subsec:tcp-kpi}
with one addition. When a retransmission timeout occurs on a subflow, its
unacknowledged segments are retransmitted both on the subflow itself, as well as
on another (they is \emph{reinjected} on another subflow). We record the
number of reinjections done by a subflow in a new \textsc{kpi} present in the
ancillary state of the \MPTCP{} subflows. Additionally, the meta-socket
provides a
bytestream service pretending to be \TCP{}. As such, it supports most of the
\KPIs{} supported by \TCP{}, with four tweaks. First, as it gets its segments
from underlying \TCP{} connections, it cannot receive corrupted segments and has
no concept of latency, removing those \KPIs{}. Second, segments arriving
out-of-order on the meta-socket no longer indicate reordering happening in the
network. Indeed, such reordering is hidden by the subflows. Instead,
reordering on the meta-socket is instead tied to the relative performance
difference between the subflows\footnote{%
    Consider two successive segments A and B, such that A comes first in the
    \MPTCP{} bytestream. If B arrives before A on the receiver's meta-socket,
    it then follows that: $(i)$ A and B were sent over different subflows, as
    subflows guarantee in-order delivery; and $(ii)$ the subflow of B was
    ``better'', e.g., had a lower latency, and/or less losses.
}. Third, duplicate incast segments now indicate reinjections.
Finally, retransmission timeouts at the meta-socket level indicate that
the connection
is suffering from head-of-line blocking (e.g., a lossy subflow prevents all
others from making progress). As one of the more common causes of such a
behaviour are too small receive buffers, this defines a new \textsc{kpi} specific
to the meta-socket.

\myitem{\eBPF{} probes handlers.} All probes defined
in~\S\ref{subsec:tcp-probes} also record the performance of
\MPTCP{} subflows as-is. In addition to them, we update the ancillary state
tracking reinjection across subflows by instrumenting
\Cword{\_\_mptcp\_reinject\_data}. Recording the performance of the meta-socket
also requires the addition of probes to record the expiration of its retransmission timer
(\Cword{mptcp\_meta\_retransmit\_timer}). New subflows initiated by the
instrumented host are automatically handled by the probes handling the creation
of \TCP{} connections. Detecting the creation of new subflows initiated by the
remote host requires instrumenting \textsc{mptcp\_check\_req\_child}.

\section{Evaluation}\label{sec:eval}

\begin{figure*}[ht]
    \hfill{}
    \def\ntttcpmkPlot#1#2{%
            \addplot+[mRed]
                    table[col sep=comma, x=value, y=proba]{%
                        figures/data/naive_data/ntttcp_cdf_#1_#2_true.csv};
            \addplot+[mBlue]
                    table[col sep=comma, x=value, y=proba]{%
                        figures/data/naive_data/ntttcp_cdf_#1_#2_false.csv};
            \addplot+[darkgreen]
                    table[col sep=comma, x=value, y=proba]{%
                        figures/data/naive_data/ntttcp_cdf_#1_#2_naive.csv};
    }
    \def\subfigWidth{.3\linewidth}\def\subfigHeight{4cm}%
    \subfloat[\sysName{} induces a small overhead when used over a link
    with no loss and sub-ms RTT.\label{fig:ntttcp_base}]{%
        \inputtikz{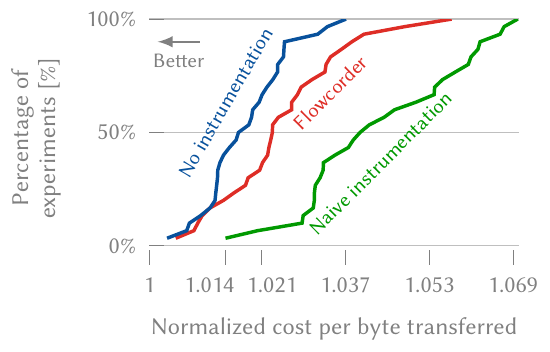}
    }\hfill{}
    \subfloat[Introducing a 10ms RTT and 0.1\% loss rate reduces the overhead by an
    order of magnitude.\label{fig:ntttcp_med}]{%
        \inputtikz{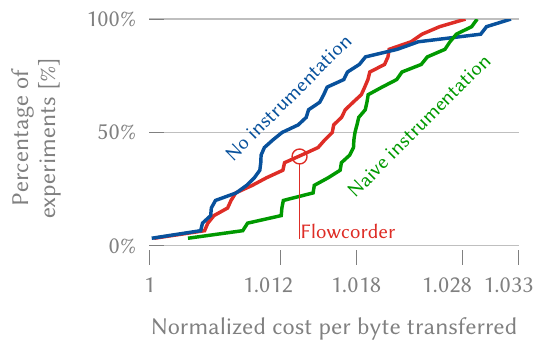}
    }\hfill{}
    \subfloat[30ms RTT and 1\% losses cause overhead of \sysName{} to become
    negligible.\label{fig:ntttcp_worst}]{%
        \inputtikz{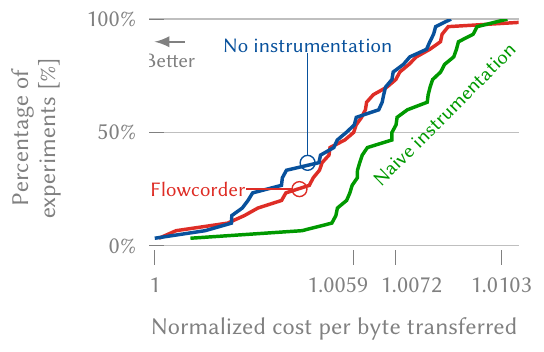}
    }\hfill{}%
    \caption{Analyzing the number of instructions executed to saturate a 10G link shows
    that the overhead induced by the kernel probes is negligible, especially
    when the link exhibits losses or reordering.}\label{fig:ntttcp}
\end{figure*}

In this section, we begin by evaluating the overhead of \system{} when
instrumenting the Linux \TCP{} stack. We first run
micro-benchmarks to estimate the overhead of \system{} in function of
on the characteristics of the underlying network~(\S\ref{subsec:overhead}).
Then, we evaluate the application-visible performance impact of instrumenting
the \TCP{} stack~(\S\ref{subsec:app_perf}). Both sets of experiments confirm
that using \system{} induces close to no performance overhead on the end-hosts.

Finally, we conclude the section by presenting how to verify that the
performance profiles produced by \system{} are accurate, especially after
kernel upgrades containing potential changes in the instrumented protocol
implementation~(\S\ref{subsec:packetdrill}). We confirme that \system{}
supports multiple versions of Linux (v4.5 to v4.18) without any modification.

\subsection{Instrumentation overhead}\label{subsec:overhead}

To estimate the overhead induced by the monitoring daemons as well as
the kernel probes injected in the \TCP{} stack by \system{}, we use a simple
benchmark between two servers (each with 8-cores\textsc{cpu}s at 2.5Ghz and 8G
of \textsc{ram}) and connected using 10G interfaces. We use
\textsc{ntttcp}~\cite{ntttcp} to initiate multiple parallel \TCP{}
connections from one server to the other (between 8 and 100), effectively
saturating the 10G link. For each experiment, we record how many bytes were
successfully transferred, and use \texttt{perf}~\cite{perf} to record the number of
\textsc{cpu} instructions that were executed during each experiment, as reported by
the hardware counters. Each experiment ran for 60 seconds, in order to average
out measurement errors. To evaluate all instrumented code paths, we also vary
the \RTT{} applied over the link (from a few hundred $\mu$s to 100ms), its
jitter (10\% of the \RTT{}), and its loss rate (from 0 to 1\% of random
losses). We performed 100 experiments per combination of \RTT{} and loss rate.

To provide quantitative baselines, we repeated each benchmark three times:
$(i)$ without any instrumentation; $(ii)$ with \system{} running on a
server; and $(iii)$ with a naive \eBPF{} instrumentation. This naive
instrumentation consists of a version of \system{} where an \eBPF{} probe
updates \KPIs{} at each incoming segment once the connection reached the
\textsc{established} state, i.e., it instruments
\Cword{tcp\_rcv\_state\_process} in place of
\Cword{tcp\_validate\_incoming} to detect out-of-order segments, or incast
duplicates. We define the instrumentation
overhead as the average number of instructions executed on the servers,
divided by the number of bytes successfully transferred. On one hand, this
metric let us easily quantify the overhead induced by \system{} as it directly
gives the amount of extra work carried by a server to execute the \eBPF{}
probes. On the other hand, we can compare the gains brought by carefully
selecting the probe locations by comparing the overhead of the two different
instrumentations. Moreover, the probe induced by the naive implementation
is executed for every incoming segment but rarely does any significant work as
few segments cause \textsc{kpi} changes (i.e., it often results in a no-op). As such, it
implicitly estimates the intrinsic overhead of placing a probe in the \TCP{}
``hot'' path (i.e., the cost of the software interrupt and the preparation of the
\eBPF{} stack). Using the number of executed \textsc{cpu}
instructions as metric has at least four advantages:
$(i)$ it is independent of the precise duration of the experiment (i.e.,
coarse-grained timers have no incidence on the results);
$(ii)$ it isolates the results from the transient states of \TCP{} congestion
control; $(iii)$ it is independent of the \textsc{cpu} frequency, which is
adjusted dynamically by the \textsc{cpu};
and $(iv)$ it captures both the load induced by the kernel probes and the load
induced by the user-space daemons aggregating \KPIs{} and exporting \IPFIX{}
records. We show a summary of the results in Fig.~\ref{fig:ntttcp}, which plots
the cumulative distribution of the fraction of experiments according to their
normalized cost (i.e., we normalize all costs by the lowest one).

When operating over a perfect link (Fig.~\ref{fig:ntttcp_base}), we see that
\system{} increases by less than 1\% the number of instructions executed during
a test. As the experiments had almost no delay and no losses, this gives a
baseline as how expensive it is to run \system{}, when all connections are
processed in the kernel fast path (i.e., the path levering as many
optimizations as possible, such as hardware offload or skb coalescing, which
decreases the overall \textsc{cpu} cost of the connection) thus
triggering as few events as possible. This contrasts with the naive
instrumentation which has an overhead of more than 2\%.
When adding some delay (10ms of \RTT{}, and 1ms of jitter), and a small random
loss probability of 0.1\%, we see in Fig.~\ref{fig:ntttcp_med} that the
per-byte instruction overhead decreases quite substantially to
approximately 0.3\%.
Indeed, as segments start to arrive out-of-order, or are
lost, the \TCP{} stack begins to process them in the slow path, which
is much more expensive \textsc{cpu}-wise than the load induced by \system{}.
This impact is even more visible as we reach a \RTT{} of 30ms$\pm$3ms,
with a loss rate of 0.5\% (Fig.~\ref{fig:ntttcp_worst}) where the overhead
induced by \system{} is almost 0.

This indicates that the relative cost of using \system{} decreases when the
network quality worsens, thus when \system{} starts to actually produce
performance profiles. The handling of lost or out-of-order segments has a much
larger impact on the performance than the kernel probes inserted by \system{}
and associated monitoring daemons. The decrease in the number of instructions
per byte transferred between Fig.~\ref{fig:ntttcp_base} and Fig.~\ref{fig:ntttcp_med}
is expected, as increasing the \RTT{} by several orders of magnitude increases
the idle periods of connections as they wait for \textsc{ack}s.

We performed the same experiments when instrumenting the \MPTCP{}
implementation~(\S\ref{subsec:mptcp}) and observed similar overhead figures,
although there were almost no differences between the two instrumentations
as \MPTCP{} disables the kernel \TCP{} fast path processing. Finally, we stress
that \system{}'s memory overhead is limited by design, as it only has to
allocate memory for the ancillary state (bounded by default to about 600kb,
i.e., 3000 \TCP{} flows), as well as a python VM holding an MTU-sized \IPFIX{}
buffer.

\subsection{Impact on application performance}\label{subsec:app_perf}

\begin{figure*}[ht]
    \def\subfigWidth{.215\linewidth}\def\subfigHeight{2.7cm}%
    \subfloat[\sysName{} has a negligible overhead on the overall completion
    time when downloading a file.\label{fig:apach_bench_base}]{%
        \inputtikz{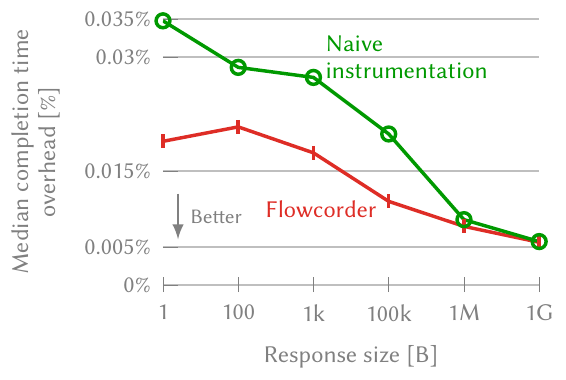}%
    }\hfill%
    \subfloat[The instrumentation overhead is comparatively higher when
    downloading one-byte files.\label{fig:apach_bench_1b}]{%
        \inputtikz{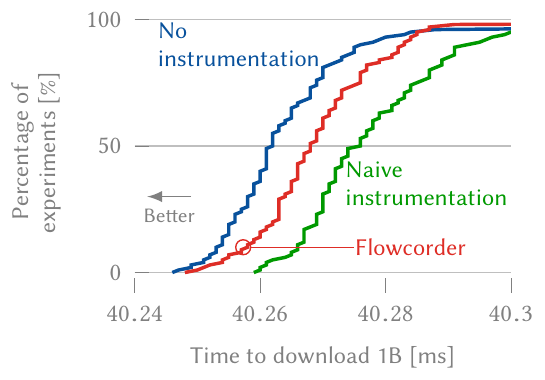}%
    }\hfill%
    \subfloat[Large transfers amortize the overhead as few performance
    profiles are generated per connection.\label{fig:apach_bench_1g}]{%
        \inputtikz{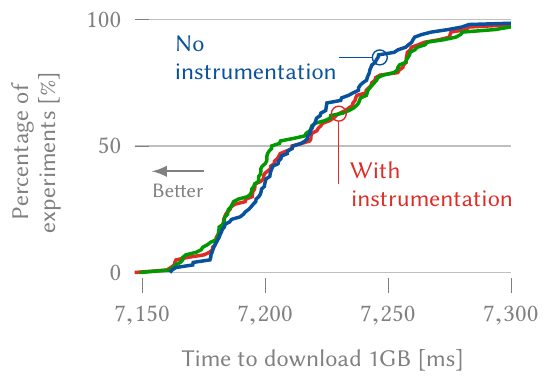}%
    }%
    \caption{\vspace{-2ex}Using \sysName{} has almost no application-visible impact on the
    performance of the Linux \TCP{} stack.}\label{fig:apache}
\end{figure*}

The previous section showed that \system{} was inducing some overhead on the
instrumented end hosts. In this section, we evaluate whether this overhead can
cause application-visible performance degradations. To this end, we configure one
host to run a \HTTP{} server. We then record the time to perform an
\HTTP{}~\textsc{get} to download a file of a given size from the
server. As we saw earlier~(\S\ref{subsec:overhead}), the overhead of \system{} is
maximum in a perfect network. As such, we directly connect both the client and
the server, configure their interfaces to induce a 20ms \RTT{}, and enable
Ethernet flow-control to prevent packet losses.
We simulate the client requests using ApacheBench~\cite{apache-bench}, with
a variable number of parallel connections (up to 100). Each experiment is
repeated 2000 times (i.e., we open a total of 2000 connections for each response
size). We recorded for each experiment how quickly the connection completed
(i.e., how long did it take to perform the TCP three-way handshake, the \HTTP{}
\textsc{get}, then download the response and close the connection).
As before, we repeated the benchmark three times (without instrument,
with \system{}, and with a naive version of \system{}). The results are visible
in Fig.~\ref{fig:apache}.

Fig.~\ref{fig:apach_bench_base} shows the median overhead per response size,
which is the observed increase in completion time when the end-host was
being instrumented by \system{}. We see that as the size of the \HTTP{} responses
increases, the overhead decreases. This result is expected. Indeed,
recall that \system{} generates at least two performance profiles for each
connection, and none in the established state if there are no performance
degradations. If the response exceeds a few \TCP{} segments, its completion time
is thus dominated by the \TCP{} data transfer, and not by the execution of
kernel probes. Fig.~\ref{fig:apach_bench_1b} thus shows the absolute worst case
for these experiments, as the response consists in a single segment. We see
that the median increase in the response time in that case is about 0.017\%.
Fig.~\ref{fig:apach_bench_1g} shows the overhead with a 1GB
response, which exhibits a much lower completion overhead. We also performed
experiments over a link with some delay and/or losses, and observed that the
overhead in those case was even lower as the response time was completely
dominated by the network characteristics.

These benchmarks, show that despite inducing some overhead, \system{}
has a very low (if not negligible) impact on the performance of connections
initiated by applications. This result also holds when instrumenting \MPTCP{}.

\subsection{Ensuring accurate measurements}\label{subsec:packetdrill}

The content of the performance profiles generated by \system{}, and thus the
accuracy of the measurements, clearly depends on the correctness of our
instrumentation of the protocol implementation.

\myitem{Sources of measurement errors.}
\system{} extracts most of its \KPIs{} by performing raw memory accesses in the
kernel's per-connection states. As the content or layout of these states could
vary across kernel versions, this extraction process is thus a first
possible source of errors. Values could be read at incorrect offsets, or be
decoded incorrectly (e.g., reading only the first 32b of a 64b counter).
A second source of possible errors are the assumptions the probes make on the
status of the connection.
For example, the \TCP{} instrumentation assumes that
a connection can be identified by the memory address at which its state
resides, which is conveniently passed around as \Cword{struct sock *sk} in most
functions. If this assumption is wrong (or no longer holds due to an update),
then \system{} will produce incorrect measurements, e.g., it might mix up
connections, or wrongly assume that a connection received an out-of-order
segment.

A third source of errors is the set of probes and their locations.
Indeed, as the implementation of the protocol improves over time, the set of
functions called for each event (e.g., received segments, timer expiration)
and their relative order might change. The most obvious effect of this on
\system{} would be inconsistent performance profiles (e.g., increasing the
number of bytes transferred of a closed connection), or missed events (e.g.,
missed \textsc{RTO}s).

\myitem{Preventing measurement errors.}
To prevent the first source of errors, \system{} re-compiles its \eBPF{} code
every time probes are inserted. As this compilation process directly
happens on the instrumented host, it can use information local to the machine
(e.g., headers matching the running kernel, or values in \texttt{procfs} to enable or
disable the \MPTCP{} instrumentation). This source of measurement errors is
thus prevented by design. Incidentally, this re-compilation process also
ensures that probes are always inserted at their proper locations, as their
offset are also dynamically computed during the \eBPF{} compilation, either by
reading the content of \Cword{/proc/kallsyms} for kernel symbols, or using
the debug symbols of user-space applications.

To prevent the seconds and third types of errors, we built a test suite using
Packetdrill~\cite{packetdrill}. Packetdrill enables us to test protocol
implementation using scripts which describe connections. More specifically,
those scripts inject crafted packets in a local interface at
specific points in time, as well as specify the content of packet(s) that
should be sent by an implementation in response to incoming packets or
\textsc{api} calls. Packetdrill contains a set of edge test cases for the
Linux \TCP{} implementation, and similar test cases for \MPTCP{} are
available~\cite{packetdrill-mptcp}. As each test case depicts a well-defined
connection, we can statically predict the performance profiles that should be
produced by \system{} when instrumenting that connection. This lets us build
integration tests to validate that \system{} accurately instruments
protocol implementations as they evolve.

Using this test suite, we were able to ensure that \system{} accurately
instruments the \TCP{} stack of the Linux kernel from v4.5 to v4.18,
and \MPTCP{} v0.93.

\section{\system{} in a campus network}\label{sec:case_study}

\def\figWidth{.24\linewidth}
\def\figHeight{2.7cm}
\begin{figure*}
    \subfloat[\vspace{-1ex}Number of \TCP{} connections in the data
        set.\label{fig:ipv46-connections}]{%
            \inputtikz{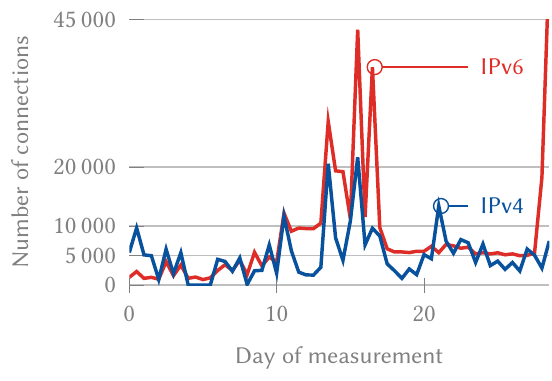}
    }%
    \hfill
    \subfloat[\vspace{-1ex}IPv4 and IPv6 have similar connection establishment
    delays, causing Happy-Eyeballs to favor IPv6.\label{fig:tcp_rtt}]{%
            \inputtikz{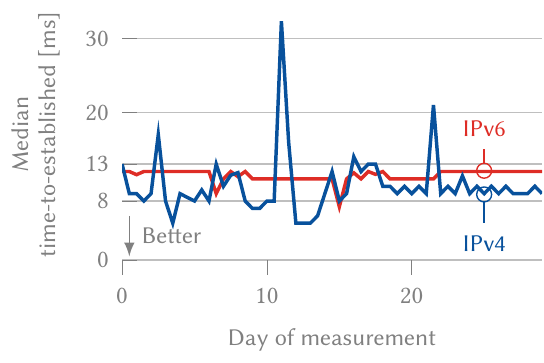}
    }%
    \hfill
    \subfloat[IPv4 connection suffer from a larger jitter.\label{fig:jitter}]{%
            \inputtikz{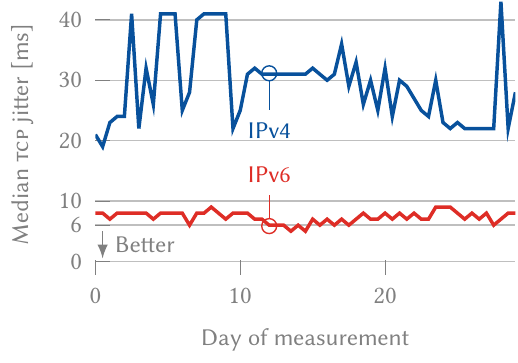}
    }

    \hfill
    \subfloat[\vspace{-1ex}A disproportionate amount of IPv4 connections
        experiences at least one initial \textsc{syn} retransmission.\label{fig:syn-retrans}]{%
            \inputtikz{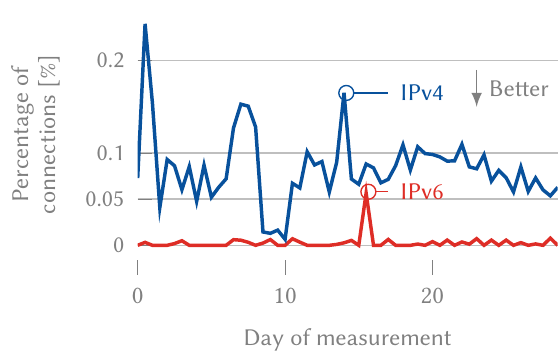}
    }
    \hfill
    \subfloat[\vspace{-1ex}One large cloud service provider often routes
        requests towards datacenters inducing large \RTT{}s.\label{fig:googlems}]{%
            \inputtikz{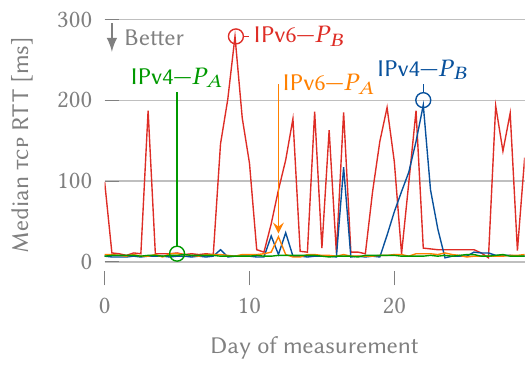}
    }
    \hfill
    \subfloat[\vspace{-1ex}An issue in one datacenter caused
        colocated service to present vastly different response time.\label{fig:tcp_dns}]{%
            \inputtikz{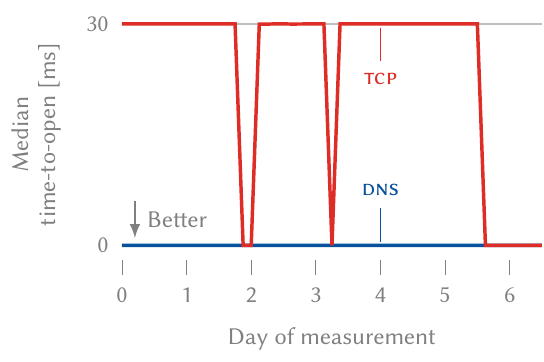}
    }
    \hfill\caption{\vspace{-2ex}Network performance insights provided by \sysName{} in a
        dual-stacked, multi-homed, campus network.}\label{fig:case_study}
\end{figure*}

We now present measurements collected over one month with \system{} in a campus network.
We deployed \system{} in student computer labs, where we run on every host
monitoring daemons that instrument the Linux kernel \TCP{} stack, presented in
\S\ref{sec:tcp}, as well as \DNS{} resolutions libraries.
Each end-host is dual-stacked and has public addresses. 

\textbf{Viewing the effects of Happy Eyeballs.}
Fig.~\ref{fig:ipv46-connections} shows the repartition of the \TCP{} connections
in function of the \textsc{ip} version used. We see that most of the
connections are established using \IPv{6}. As major cloud services are very popular amongst
students and they all support \IPv{6}, this could be due to Happy
Eyeballs~\cite{happy-eyeballs}. We can confirm that Happy Eyeballs indeed
favors connections over \IPv{6} by looking at Fig.~\ref{fig:tcp_rtt}. It
compares the median time required to establish new \TCP{} connections depending on
the used address family. More specifically, it only contains connections
established towards dual-stacked \ASes{}. We see that the time to open
a new connection is similar for both address families, despite \IPv{4} exhibiting
many outliers. As Happy Eyeballs gives \IPv{6} connections a head start of
usually 300ms (although some have called to reduce
it~\cite{bajpai2016measuring}), this explains why \IPv{6} is almost
always used to reach popular services.

\textbf{Comparing the performance of different uplinks.}
Our network is dual-homed. It uses different uplinks for \IPv{4} and 
\IPv{6}. We leverage \system{} to analyze the difference between the two address families. Fig.~\ref{fig:jitter} shows the median jitter
observed for \TCP{} connections.
We observe that the jitter experienced by \IPv{4} connections is higher than
for \IPv{6}. This correlates with the trend from Fig.~\ref{fig:tcp_rtt}, where
\IPv{4} showed more variations. Finally, to better understand why the \IPv{4}
connection establishment delay had a higher variance,
Fig.~\ref{fig:syn-retrans} shows
the ratio of connections that were successfully established after losing their
initial \TCP{} \textsc{syn}. We see that this mainly occurs only for \IPv{4},
which might point to an on-site issue with a firewall or congestion of the
\IPv{4} uplink. Overall, these results show that \IPv{6}
connections seem to perform better than \IPv{4} connections in our campus.
This is expected, as only the \IPv{4} traffic is shaped by our provider.

\textbf{Comparing the performance of remote cloud services.}
Another usage for the measurements collected by \system{} is to compare the
performance when accessing different cloud services. Indeed, as an \textsc{isp}
might have different peering agreements with them, measuring
the quality of the connections towards those service can be a factor to decide
whether to subscribe to one service or another (or to select a different
\textsc{ISP}). For example,
Fig.~\ref{fig:googlems} compares the median \TCP{} \RTT{} when accessing
two popular cloud services. For these services, a low \RTT{} is key to ensure a proper
level of interactivity. We see that while both services tend to show similar
\RTT{}'s over \IPv{4}, one of them (P$_B$) performs much
worse when accessed over \IPv{6}\footnote{%
    Further analyzes revealed that the provider's \DNS{} was causing students'
    requests to use datacenters located on another continent.
}. Keep in mind that while \system{} uses
\TCP{}'s estimates to report \RTT{} and jitter, this might not completely
reflect the true values to reach the actual server, as there could be
middleboxes or \TCP{} proxies present on the path, fiddling with segments.

\textbf{Detecting a local operational issue.}
Beside providing external connectivity, our campus network also hosts services
such as a \DNS{} resolver or institutional web servers. During our
measurement campaign, students were complaining that accessing those
web servers was abnormally slow. As these web servers are
collocated with the \DNS{} servers, we can thus directly use \system{}
to compare their
performance. Fig.~\ref{fig:tcp_dns} shows the median time to establish a
connection to any of these servers. Given that the servers are
located a few hundreds of meters away, 30ms to
receive a \textsc{syn+ack} is a clear performance anomaly, especially compared
to the time required to receive a \DNS{} reply. After talking with the network
operators, we learned that this problem was due to a faulty load-balancer that was
fixed near the end
of the observation period.

\section{Related work}

Monitoring network performance is an age-old topic.
\system{} draws from three main threads of work.

\textbf{Collecting transport performance metrics.}
Passive inference of transport protocol characteristics has been a primary
source of measurements for a long time, e.g., inferring per-flow \TCP{} states
by analyzing packet headers provided by a network tap
(tstat~\citet{Mellia_tstat:2002}), or correlating packet traces collected on
the end hosts (Deja-vu~\cite{deja-vu}). More recent approaches tailored to
data-centers (e.g., Trumpet~\cite{trumpet}, Dapper~\cite{dapper17}) perform
such analyzes in real-time, at the edges of the network (i.e., access switches
or virtual machine hypervisors). While these technique provide fine-grained
measurements for \TCP{} they will not be applicable to emerging encrypted
protocols such as \QUIC{}.

\textbf{Instrumenting the end-hosts.}
SNAP~\cite{snap11} or NetPoirot~\cite{netpoirot} collect an enormous amount of
statistics about \TCP{} connections directly from datacenter hosts.
By collecting those on a central management system, they can then correlate
observations in order to identify the root causes of performance issues (e.g.,
bottleneck switch or link, or misconfigured of \TCP{} delayed \textsc{ack}'s).
Both tools poll event loggers (e.g., Windows EWT, or
Linux syslog) every few milliseconds. As such, they are restricted to the
measurements provided by those loggers (typically the \TCP{}
\textsc{mib}~\cite{tcp-mib}), with a higher \textsc{cpu} overhead than
\system{}.
Odin~\cite{nsdi18-odin} is a framework injecting
javascript when serving client requests from \textsc{cdn} to perform active
measurements. While this approach collects performance metrics as
experienced by end-hosts, the measurements that it can records are, by
design, much more limited.

\textbf{Instrumenting protocol implementations.} Several tools provide some
visibility over the internals of the Linux \TCP{} stack.
tcpprobe~\cite{tcpprobe} is a kernel module which logs the evolution of the
congestion control variable in response to incoming \TCP{} segments.
tcp-tracer~\cite{tcp-tracer} reports the \TCP{} state changes (e.g.,
\textsc{new}$\rightarrow$\textsc{established}) for all connections.
\textsc{bcc}~\cite{bcc} provides several small tools, enabling
to log some aspects of \TCP{} connections. All of these tools use the same
primitives to instrument the \TCP{} stack (i.e., kprobes, often combined
with \eBPF{} handlers), but they are not coupled with entreprise management
systems.

\section{Conclusion}
\system is a new monitoring framework which directly extracts
Key Performance Indicators from the end-hosts, at specific moments in a
connection life-cycle. \system{} seamlessly integrates with existing Network Management Systems as
it generates \IPFIX{} performance profiles. Furthermore, it is
future-proof as it readily supports multipath protocols and will also
be useable with emerging encrypted protocols. \system{} has almost no runtime
overhead, and its measurement can easily be analyzed. One future research
direction would be to use the performance profiles generated by \system{} to
drive tight-control loops on network controllers, to optimize the content of
\DNS{} replies (e.g. dynamically preferring the best address family) or to
select the best performing provider in multihoming scenarios.


\subsection*{Software artefacts}
We release the sources of \system{} at
\url{https://github.com/oliviertilmans/flowcorder} under a permissive license.
These sources are primarily
composed of python (${\sim}3300$ lines) and restricted \textsc{c} that compiles to \eBPF{}
(${\sim}1900$ lines). These include the \TCP{} monitoring daemon which has been tested to
work on the Linux kernel from v4.5 to v4.18, its extension to support
\MPTCP{} v0.93, the \DNS{} monitoring daemon, and scripts to package and deploy
them. We also provide a sample \IPFIX{} collector based on an
\textsc{elk} stack~\cite{ELK} which comes with preloaded normalization filters.
Finally, to ensure the reproducibility of our results, we also provide all
scripts used to conduct the benchmarks reported in \S\ref{sec:eval}.

\clearpage
{
    \small
    \bibliographystyle{ACM-Reference-Format}
    \nocite{*}
    \bibliography{biblio}


\begin{thebibliography}{84}


\ifx \showCODEN    \undefined \def \showCODEN     #1{\unskip}     \fi
\ifx \showDOI      \undefined \def \showDOI       #1{#1}\fi
\ifx \showISBNx    \undefined \def \showISBNx     #1{\unskip}     \fi
\ifx \showISBNxiii \undefined \def \showISBNxiii  #1{\unskip}     \fi
\ifx \showISSN     \undefined \def \showISSN      #1{\unskip}     \fi
\ifx \showLCCN     \undefined \def \showLCCN      #1{\unskip}     \fi
\ifx \shownote     \undefined \def \shownote      #1{#1}          \fi
\ifx \showarticletitle \undefined \def \showarticletitle #1{#1}   \fi
\ifx \showURL      \undefined \def \showURL       {\relax}        \fi
\providecommand\bibfield[2]{#2}
\providecommand\bibinfo[2]{#2}
\providecommand\natexlab[1]{#1}
\providecommand\showeprint[2][]{arXiv:#2}

\bibitem[\protect\citeauthoryear{Aggarwal, Bhagwan, De~Carli, Padmanabhan, and
  Puttaswamy}{Aggarwal et~al\mbox{.}}{2011}]%
        {deja-vu}
\bibfield{author}{\bibinfo{person}{Bhavish Aggarwal}, \bibinfo{person}{Ranjita
  Bhagwan}, \bibinfo{person}{Lorenzo De~Carli}, \bibinfo{person}{Venkat
  Padmanabhan}, {and} \bibinfo{person}{Krishna Puttaswamy}.}
  \bibinfo{year}{2011}\natexlab{}.
\newblock \showarticletitle{Deja Vu: Fingerprinting Network Problems}. In
  \bibinfo{booktitle}{\emph{CoNEXT '11}}. \bibinfo{publisher}{ACM},
  \bibinfo{address}{New York, NY, USA}, Article \bibinfo{articleno}{28},
  \bibinfo{numpages}{12}~pages.
\newblock
\showISBNx{978-1-4503-1041-3}
\urldef\tempurl%
\url{https://doi.org/10.1145/2079296.2079324}
\showDOI{\tempurl}


\bibitem[\protect\citeauthoryear{Anastasov}{Anastasov}{[n. d.]}]%
        {tcpmetrics}
\bibfield{author}{\bibinfo{person}{Julian Anastasov}.} \bibinfo{year}{[n.
  d.]}\natexlab{}.
\newblock \bibinfo{title}{ip-tcp\_metrics - management for TCP Metrics}.
\newblock
\newblock
\urldef\tempurl%
\url{https://www.linux.org/docs/man8/ip-tcp_metrics.html}
\showURL{%
Retrieved 2018 from \tempurl}


\bibitem[\protect\citeauthoryear{Apple}{Apple}{[n. d.]}]%
        {ios-mptcp}
\bibfield{author}{\bibinfo{person}{Apple}.} \bibinfo{year}{[n. d.]}\natexlab{}.
\newblock \bibinfo{title}{Advances in Networking}.
\newblock
\newblock
\urldef\tempurl%
\url{https://developer.apple.com/videos/play/wwdc2017/707/}
\showURL{%
Retrieved 2018 from \tempurl}


\bibitem[\protect\citeauthoryear{Arzani, Ciraci, Loo, Schuster, and
  Outhred}{Arzani et~al\mbox{.}}{2016}]%
        {netpoirot}
\bibfield{author}{\bibinfo{person}{Behnaz Arzani}, \bibinfo{person}{Selim
  Ciraci}, \bibinfo{person}{Boon~Thau Loo}, \bibinfo{person}{Assaf Schuster},
  {and} \bibinfo{person}{Geoff Outhred}.} \bibinfo{year}{2016}\natexlab{}.
\newblock \showarticletitle{Taking the Blame Game out of Data Centers
  Operations with NetPoirot}. In \bibinfo{booktitle}{\emph{SIGCOMM '16}}.
  \bibinfo{publisher}{ACM}, \bibinfo{address}{New York, NY, USA},
  \bibinfo{pages}{440--453}.
\newblock
\showISBNx{978-1-4503-4193-6}
\urldef\tempurl%
\url{https://doi.org/10.1145/2934872.2934884}
\showDOI{\tempurl}


\bibitem[\protect\citeauthoryear{Augustin, Friedman, and Teixeira}{Augustin
  et~al\mbox{.}}{2011}]%
        {augustin2011measuring}
\bibfield{author}{\bibinfo{person}{Brice Augustin}, \bibinfo{person}{Timur
  Friedman}, {and} \bibinfo{person}{Renata Teixeira}.}
  \bibinfo{year}{2011}\natexlab{}.
\newblock \showarticletitle{Measuring multipath routing in the internet}.
\newblock \bibinfo{journal}{\emph{IEEE/ACM Transactions on Networking (TON)}}
  \bibinfo{volume}{19}, \bibinfo{number}{3} (\bibinfo{year}{2011}),
  \bibinfo{pages}{830--840}.
\newblock


\bibitem[\protect\citeauthoryear{authors}{authors}{[n. d.]}]%
        {dtrace}
\bibfield{author}{\bibinfo{person}{Dtrace authors}.} \bibinfo{year}{[n.
  d.]}\natexlab{}.
\newblock \bibinfo{title}{About DTrace}.
\newblock
\newblock
\urldef\tempurl%
\url{dtrace.org}
\showURL{%
Retrieved 2018 from \tempurl}


\bibitem[\protect\citeauthoryear{Bajpai and Sch{\"o}nw{\"a}lder}{Bajpai and
  Sch{\"o}nw{\"a}lder}{2016}]%
        {bajpai2016measuring}
\bibfield{author}{\bibinfo{person}{Vaibhav Bajpai} {and}
  \bibinfo{person}{J{\"u}rgen Sch{\"o}nw{\"a}lder}.}
  \bibinfo{year}{2016}\natexlab{}.
\newblock \showarticletitle{Measuring the effects of happy eyeballs}. In
  \bibinfo{booktitle}{\emph{Proceedings of the 2016 Applied Networking Research
  Workshop}}. ACM, \bibinfo{pages}{38--44}.
\newblock


\bibitem[\protect\citeauthoryear{Bellardo and Savage}{Bellardo and
  Savage}{2002}]%
        {bellardo2002measuring}
\bibfield{author}{\bibinfo{person}{John Bellardo} {and} \bibinfo{person}{Stefan
  Savage}.} \bibinfo{year}{2002}\natexlab{}.
\newblock \showarticletitle{Measuring packet reordering}. In
  \bibinfo{booktitle}{\emph{Proceedings of the 2nd ACM SIGCOMM Workshop on
  Internet measurment}}. ACM, \bibinfo{pages}{97--105}.
\newblock


\bibitem[\protect\citeauthoryear{Bonaventure and Seo}{Bonaventure and
  Seo}{2016}]%
        {bonaventure2016multipath}
\bibfield{author}{\bibinfo{person}{O. Bonaventure} {and} \bibinfo{person}{S.
  Seo}.} \bibinfo{year}{2016}\natexlab{}.
\newblock \showarticletitle{Multipath TCP deployments}.
\newblock \bibinfo{journal}{\emph{IETF Journal}} \bibinfo{volume}{12},
  \bibinfo{number}{2} (\bibinfo{year}{2016}), \bibinfo{pages}{24--27}.
\newblock


\bibitem[\protect\citeauthoryear{Boschi, Trammell, Mark, and Zseby}{Boschi
  et~al\mbox{.}}{2009}]%
        {ipfix-custom-ie}
\bibfield{author}{\bibinfo{person}{E. Boschi}, \bibinfo{person}{B. Trammell},
  \bibinfo{person}{L. Mark}, {and} \bibinfo{person}{T. Zseby}.}
  \bibinfo{year}{2009}\natexlab{}.
\newblock \bibinfo{booktitle}{\emph{Exporting Type Information for IP Flow
  Information Export (IPFIX) Information Elements}}.
\newblock \bibinfo{type}{RFC} 5610. \bibinfo{institution}{RFC Editor}.
\newblock
\showISSN{2070-1721}


\bibitem[\protect\citeauthoryear{Calder, Schr{\"o}der, Ryan~Gao, Padhye,
  Mahajan, Ananthanarayanan, and Katz-Bassett}{Calder et~al\mbox{.}}{2018}]%
        {nsdi18-odin}
\bibfield{author}{\bibinfo{person}{Matt Calder}, \bibinfo{person}{Manuel
  Schr{\"o}der}, \bibinfo{person}{Ryan~Stewart Ryan~Gao},
  \bibinfo{person}{Jitendra Padhye}, \bibinfo{person}{Ratul Mahajan},
  \bibinfo{person}{Ganesh Ananthanarayanan}, {and} \bibinfo{person}{Ethan
  Katz-Bassett}.} \bibinfo{year}{2018}\natexlab{}.
\newblock \showarticletitle{Odin: Microsoft’s Scalable Fault-Tolerant CDN
  Measurement System}. In \bibinfo{booktitle}{\emph{NSDI 18}}. USENIX
  Association.
\newblock


\bibitem[\protect\citeauthoryear{Cardwell, Cheng, Brakmo, Mathis, Raghavan,
  Dukkipati, Chu, Terzis, and Herbert}{Cardwell et~al\mbox{.}}{2013}]%
        {packetdrill}
\bibfield{author}{\bibinfo{person}{Neal Cardwell}, \bibinfo{person}{Yuchung
  Cheng}, \bibinfo{person}{Lawrence Brakmo}, \bibinfo{person}{Matt Mathis},
  \bibinfo{person}{Barath Raghavan}, \bibinfo{person}{Nandita Dukkipati},
  \bibinfo{person}{Hsiao-keng~Jerry Chu}, \bibinfo{person}{Andreas Terzis},
  {and} \bibinfo{person}{Tom Herbert}.} \bibinfo{year}{2013}\natexlab{}.
\newblock \showarticletitle{packetdrill: Scriptable Network Stack Testing, from
  Sockets to Packets.}. In \bibinfo{booktitle}{\emph{USENIX Annual Technical
  Conference}}. \bibinfo{pages}{213--218}.
\newblock


\bibitem[\protect\citeauthoryear{Cardwell, Cheng, Gunn, Yeganeh, and
  Jacobson}{Cardwell et~al\mbox{.}}{2016}]%
        {tcp-bbr}
\bibfield{author}{\bibinfo{person}{Neal Cardwell}, \bibinfo{person}{Yuchung
  Cheng}, \bibinfo{person}{C~Stephen Gunn}, \bibinfo{person}{Soheil~Hassas
  Yeganeh}, {and} \bibinfo{person}{Van Jacobson}.}
  \bibinfo{year}{2016}\natexlab{}.
\newblock \showarticletitle{BBR: Congestion-based congestion control}.
\newblock \bibinfo{journal}{\emph{Queue}} \bibinfo{volume}{14},
  \bibinfo{number}{5} (\bibinfo{year}{2016}), \bibinfo{pages}{50}.
\newblock


\bibitem[\protect\citeauthoryear{Case, Fedor, Schoffstall, and Davin}{Case
  et~al\mbox{.}}{1990}]%
        {snmp}
\bibfield{author}{\bibinfo{person}{Jeffrey~D. Case}, \bibinfo{person}{Mark
  Fedor}, \bibinfo{person}{Martin~Lee Schoffstall}, {and}
  \bibinfo{person}{James~R. Davin}.} \bibinfo{year}{1990}\natexlab{}.
\newblock \bibinfo{booktitle}{\emph{Simple Network Management Protocol
  (SNMP)}}.
\newblock \bibinfo{type}{STD}~15.
\newblock
\showISSN{2070-1721}


\bibitem[\protect\citeauthoryear{Chung, Choffnes, and Mislove}{Chung
  et~al\mbox{.}}{2016}]%
        {chung2016tunneling}
\bibfield{author}{\bibinfo{person}{Taejoong Chung}, \bibinfo{person}{David
  Choffnes}, {and} \bibinfo{person}{Alan Mislove}.}
  \bibinfo{year}{2016}\natexlab{}.
\newblock \showarticletitle{Tunneling for transparency: A large-scale analysis
  of end-to-end violations in the internet}. In
  \bibinfo{booktitle}{\emph{IMC'16}}. ACM, \bibinfo{pages}{199--213}.
\newblock


\bibitem[\protect\citeauthoryear{Cisco}{Cisco}{[n. d.]}]%
        {ipsla}
\bibfield{author}{\bibinfo{person}{Cisco}.} \bibinfo{year}{[n. d.]}\natexlab{}.
\newblock \bibinfo{title}{IP SLAs Configuration Guide}.
\newblock
\newblock
\urldef\tempurl%
\url{https://www.cisco.com/c/en/us/td/docs/ios-xml/ios/ipsla/configuration/xe-16/sla-xe-16-book.html}
\showURL{%
Retrieved 2018 from \tempurl}


\bibitem[\protect\citeauthoryear{Cloudflare}{Cloudflare}{[n. d.]}]%
        {cloudflare-dns}
\bibfield{author}{\bibinfo{person}{Cloudflare}.} \bibinfo{year}{[n.
  d.]}\natexlab{}.
\newblock \bibinfo{title}{1.1.1.1 -- the Internet's fastest, privacy-first DNS
  resolver}.
\newblock
\newblock
\urldef\tempurl%
\url{https://1.1.1.1/}
\showURL{%
\tempurl}


\bibitem[\protect\citeauthoryear{Corbet}{Corbet}{2007}]%
        {gro}
\bibfield{author}{\bibinfo{person}{Jonathan Corbet}.}
  \bibinfo{year}{2007}\natexlab{}.
\newblock \bibinfo{title}{Large receive offload}.
\newblock
\newblock
\urldef\tempurl%
\url{https://lwn.net/Articles/243949/}
\showURL{%
Retrieved 2018 from \tempurl}


\bibitem[\protect\citeauthoryear{Crocker and Overell}{Crocker and
  Overell}{2008}]%
        {rfc5234}
\bibfield{author}{\bibinfo{person}{D. Crocker} {and} \bibinfo{person}{P.
  Overell}.} \bibinfo{year}{2008}\natexlab{}.
\newblock \bibinfo{booktitle}{\emph{Augmented BNF for Syntax Specifications:
  ABNF}}.
\newblock \bibinfo{type}{STD}~68. \bibinfo{institution}{RFC Editor}.
\newblock
\showISSN{2070-1721}


\bibitem[\protect\citeauthoryear{De~Coninck and Bonaventure}{De~Coninck and
  Bonaventure}{2017}]%
        {de2017multipath}
\bibfield{author}{\bibinfo{person}{Quentin De~Coninck} {and}
  \bibinfo{person}{Olivier Bonaventure}.} \bibinfo{year}{2017}\natexlab{}.
\newblock \showarticletitle{Multipath QUIC: Design and Evaluation}. In
  \bibinfo{booktitle}{\emph{Conext'17}}. ACM, \bibinfo{pages}{160--166}.
\newblock


\bibitem[\protect\citeauthoryear{De~Vaere, B{\"u}hler, K{\"u}hlewind, and
  Trammell}{De~Vaere et~al\mbox{.}}{2018}]%
        {de2018three}
\bibfield{author}{\bibinfo{person}{Piet De~Vaere}, \bibinfo{person}{Tobias
  B{\"u}hler}, \bibinfo{person}{Mirja K{\"u}hlewind}, {and}
  \bibinfo{person}{Brian Trammell}.} \bibinfo{year}{2018}\natexlab{}.
\newblock \showarticletitle{Three Bits Suffice: Explicit Support for Passive
  Measurement of Internet Latency in QUIC and TCP}. In
  \bibinfo{booktitle}{\emph{IMC'18}}.
\newblock


\bibitem[\protect\citeauthoryear{Drago, Hofstede, Sadre, Sperotto, and
  Pras}{Drago et~al\mbox{.}}{2015}]%
        {drago2015measuring}
\bibfield{author}{\bibinfo{person}{Idilio Drago}, \bibinfo{person}{Rick
  Hofstede}, \bibinfo{person}{Ramin Sadre}, \bibinfo{person}{Anna Sperotto},
  {and} \bibinfo{person}{Aiko Pras}.} \bibinfo{year}{2015}\natexlab{}.
\newblock \showarticletitle{Measuring cloud service health using NetFlow/IPFIX:
  the WikiLeaks case}.
\newblock \bibinfo{journal}{\emph{Journal of network and systems management}}
  \bibinfo{volume}{23}, \bibinfo{number}{1} (\bibinfo{year}{2015}),
  \bibinfo{pages}{58--88}.
\newblock


\bibitem[\protect\citeauthoryear{Duchene and Bonaventure}{Duchene and
  Bonaventure}{2017}]%
        {duchene-mptcp-lb}
\bibfield{author}{\bibinfo{person}{Fabien Duchene} {and}
  \bibinfo{person}{Olivier Bonaventure}.} \bibinfo{year}{2017}\natexlab{}.
\newblock \showarticletitle{Making multipath TCP friendlier to load balancers
  and anycast}. In \bibinfo{booktitle}{\emph{2017 IEEE 25th International
  Conference on Network Protocols (ICNP)}}. IEEE, \bibinfo{pages}{1--10}.
\newblock


\bibitem[\protect\citeauthoryear{Elasticsearch}{Elasticsearch}{[n. d.]}]%
        {ELK}
\bibfield{author}{\bibinfo{person}{Elasticsearch}.} \bibinfo{year}{[n.
  d.]}\natexlab{}.
\newblock \bibinfo{title}{ELK Stack: Elasticsearch, Logstash, Kibana}.
\newblock
\newblock
\urldef\tempurl%
\url{https://www.elastic.co/elk-stack}
\showURL{%
Retrieved 2018 from \tempurl}


\bibitem[\protect\citeauthoryear{et~al.}{et~al.}{[n. d.]}]%
        {tcpcrypt}
\bibfield{author}{\bibinfo{person}{Andrea~Bittau et al.}} \bibinfo{year}{[n.
  d.]}\natexlab{}.
\newblock \bibinfo{title}{Tcpcrypt -- Encrypting the Internet}.
\newblock
\newblock
\urldef\tempurl%
\url{http://tcpcrypt.org/}
\showURL{%
Retrieved 2018 from \tempurl}


\bibitem[\protect\citeauthoryear{Farrell and Tschofenig}{Farrell and
  Tschofenig}{2014}]%
        {rfc7258}
\bibfield{author}{\bibinfo{person}{S. Farrell} {and} \bibinfo{person}{H.
  Tschofenig}.} \bibinfo{year}{2014}\natexlab{}.
\newblock \bibinfo{booktitle}{\emph{Pervasive Monitoring Is an Attack}}.
\newblock \bibinfo{type}{BCP} 188.
\newblock
\showISSN{2070-1721}


\bibitem[\protect\citeauthoryear{Felt, Barnes, King, Palmer, Bentzel, and
  Tabriz}{Felt et~al\mbox{.}}{2017}]%
        {felt2017measuring}
\bibfield{author}{\bibinfo{person}{Adrienne~Porter Felt},
  \bibinfo{person}{Richard Barnes}, \bibinfo{person}{April King},
  \bibinfo{person}{Chris Palmer}, \bibinfo{person}{Chris Bentzel}, {and}
  \bibinfo{person}{Parisa Tabriz}.} \bibinfo{year}{2017}\natexlab{}.
\newblock \showarticletitle{Measuring HTTPS adoption on the web}. In
  \bibinfo{booktitle}{\emph{26th USENIX Security Symposium}}.
  \bibinfo{pages}{1323--1338}.
\newblock


\bibitem[\protect\citeauthoryear{Finamore, Mellia, Meo, Munafo, Di~Torino, and
  Rossi}{Finamore et~al\mbox{.}}{2011}]%
        {finamore2011experiences}
\bibfield{author}{\bibinfo{person}{Alessandro Finamore}, \bibinfo{person}{Marco
  Mellia}, \bibinfo{person}{Michela Meo}, \bibinfo{person}{Maurizio~M Munafo},
  \bibinfo{person}{Politecnico Di~Torino}, {and} \bibinfo{person}{Dario
  Rossi}.} \bibinfo{year}{2011}\natexlab{}.
\newblock \showarticletitle{Experiences of internet traffic monitoring with
  tstat}.
\newblock \bibinfo{journal}{\emph{IEEE Network}} \bibinfo{volume}{25},
  \bibinfo{number}{3} (\bibinfo{year}{2011}), \bibinfo{pages}{8--14}.
\newblock


\bibitem[\protect\citeauthoryear{Ford, Raiciu, Handley, and Bonaventure}{Ford
  et~al\mbox{.}}{2013}]%
        {RFC6824}
\bibfield{author}{\bibinfo{person}{A. Ford}, \bibinfo{person}{C. Raiciu},
  \bibinfo{person}{M. Handley}, {and} \bibinfo{person}{O. Bonaventure}.}
  \bibinfo{year}{2013}\natexlab{}.
\newblock \bibinfo{booktitle}{\emph{TCP Extensions for Multipath Operation with
  Multiple Addresses}}.
\newblock \bibinfo{type}{RFC} 6824.
\newblock
\showISSN{2070-1721}


\bibitem[\protect\citeauthoryear{Foundation}{Foundation}{[n. d.]}]%
        {apache-bench}
\bibfield{author}{\bibinfo{person}{The Apache~Software Foundation}.}
  \bibinfo{year}{[n. d.]}\natexlab{}.
\newblock \bibinfo{title}{ab - Apache HTTP server benchmarking tool}.
\newblock
\newblock
\urldef\tempurl%
\url{https://httpd.apache.org/docs/2.4/programs/ab.html}
\showURL{%
Retrieved 2018 from \tempurl}


\bibitem[\protect\citeauthoryear{Gangam, Sharma, and Fahmy}{Gangam
  et~al\mbox{.}}{2013}]%
        {gangam2013pegasus}
\bibfield{author}{\bibinfo{person}{Sriharsha Gangam}, \bibinfo{person}{Puneet
  Sharma}, {and} \bibinfo{person}{Sonia Fahmy}.}
  \bibinfo{year}{2013}\natexlab{}.
\newblock \showarticletitle{Pegasus: Precision hunting for icebergs and
  anomalies in network flows}. In \bibinfo{booktitle}{\emph{INFOCOM, 2013
  Proceedings IEEE}}. IEEE, \bibinfo{pages}{1420--1428}.
\newblock


\bibitem[\protect\citeauthoryear{Garcia-Dorado, Finamore, Mellia, Meo, and
  Munafo}{Garcia-Dorado et~al\mbox{.}}{2012}]%
        {Garcia_Dorado:Characterization:2012}
\bibfield{author}{\bibinfo{person}{Jose Garcia-Dorado},
  \bibinfo{person}{Alessandro Finamore}, \bibinfo{person}{Marco Mellia},
  \bibinfo{person}{Michela Meo}, {and} \bibinfo{person}{Maurizio~M. Munafo}.}
  \bibinfo{year}{2012}\natexlab{}.
\newblock \showarticletitle{Characterization of ISP Traffic: Trends, User
  Habits, and Access Technology Impact}.
\newblock \bibinfo{journal}{\emph{IEEE Transactions on Network and Service
  Management}} \bibinfo{volume}{9}, \bibinfo{number}{2} (\bibinfo{date}{Feb}
  \bibinfo{year}{2012}).
\newblock


\bibitem[\protect\citeauthoryear{Ghasemi, Benson, and Rexford}{Ghasemi
  et~al\mbox{.}}{2017}]%
        {dapper17}
\bibfield{author}{\bibinfo{person}{Mojgan Ghasemi}, \bibinfo{person}{Theophilus
  Benson}, {and} \bibinfo{person}{Jennifer Rexford}.}
  \bibinfo{year}{2017}\natexlab{}.
\newblock \showarticletitle{Dapper: Data plane performance diagnosis of tcp}.
  In \bibinfo{booktitle}{\emph{Proceedings of the Symposium on SDN Research}}.
  ACM, \bibinfo{pages}{61--74}.
\newblock


\bibitem[\protect\citeauthoryear{Google}{Google}{[n. d.]}]%
        {quic-trace}
\bibfield{author}{\bibinfo{person}{Google}.} \bibinfo{year}{[n.
  d.]}\natexlab{}.
\newblock \bibinfo{title}{QUIC Trace}.
\newblock
\newblock
\urldef\tempurl%
\url{https://github.com/google/quic-trace}
\showURL{%
Retrieved 2018 from \tempurl}


\bibitem[\protect\citeauthoryear{Goswami}{Goswami}{2005}]%
        {kprobe-lwn}
\bibfield{author}{\bibinfo{person}{Sudhanshu Goswami}.} \bibinfo{year}{April
  18, 2005}\natexlab{}.
\newblock \bibinfo{title}{An introduction to KProbes}.
\newblock
\newblock
\urldef\tempurl%
\url{https://lwn.net/Articles/132196/}
\showURL{%
\tempurl}


\bibitem[\protect\citeauthoryear{Gunnar, Johansson, and Telkamp}{Gunnar
  et~al\mbox{.}}{2004}]%
        {gunnar2004traffic}
\bibfield{author}{\bibinfo{person}{Anders Gunnar}, \bibinfo{person}{Mikael
  Johansson}, {and} \bibinfo{person}{Thomas Telkamp}.}
  \bibinfo{year}{2004}\natexlab{}.
\newblock \showarticletitle{Traffic matrix estimation on a large IP backbone: a
  comparison on real data}. In \bibinfo{booktitle}{\emph{Proceedings of the 4th
  ACM SIGCOMM conference on Internet measurement}}. ACM,
  \bibinfo{pages}{149--160}.
\newblock


\bibitem[\protect\citeauthoryear{Guo, Yuan, Xiang, Dang, Huang, Maltz, Liu,
  Wang, Pang, Chen, Lin, and Kurien}{Guo et~al\mbox{.}}{2015}]%
        {pingmesh}
\bibfield{author}{\bibinfo{person}{Chuanxiong Guo}, \bibinfo{person}{Lihua
  Yuan}, \bibinfo{person}{Dong Xiang}, \bibinfo{person}{Yingnong Dang},
  \bibinfo{person}{Ray Huang}, \bibinfo{person}{Dave Maltz},
  \bibinfo{person}{Zhaoyi Liu}, \bibinfo{person}{Vin Wang},
  \bibinfo{person}{Bin Pang}, \bibinfo{person}{Hua Chen},
  \bibinfo{person}{Zhi-Wei Lin}, {and} \bibinfo{person}{Varugis Kurien}.}
  \bibinfo{year}{2015}\natexlab{}.
\newblock \showarticletitle{Pingmesh: A Large-Scale System for Data Center
  Network Latency Measurement and Analysis}. In
  \bibinfo{booktitle}{\emph{SIGCOMM '15}}. \bibinfo{publisher}{ACM},
  \bibinfo{address}{New York, NY, USA}, \bibinfo{pages}{139--152}.
\newblock
\showISBNx{978-1-4503-3542-3}
\urldef\tempurl%
\url{https://doi.org/10.1145/2785956.2787496}
\showDOI{\tempurl}


\bibitem[\protect\citeauthoryear{Hofstede, {\v{C}}eleda, Trammell, Drago,
  Sadre, Sperotto, and Pras}{Hofstede et~al\mbox{.}}{2014}]%
        {hofstede2014flow}
\bibfield{author}{\bibinfo{person}{Rick Hofstede}, \bibinfo{person}{Pavel
  {\v{C}}eleda}, \bibinfo{person}{Brian Trammell}, \bibinfo{person}{Idilio
  Drago}, \bibinfo{person}{Ramin Sadre}, \bibinfo{person}{Anna Sperotto}, {and}
  \bibinfo{person}{Aiko Pras}.} \bibinfo{year}{2014}\natexlab{}.
\newblock \showarticletitle{Flow monitoring explained: From packet capture to
  data analysis with netflow and ipfix}.
\newblock \bibinfo{journal}{\emph{IEEE Communications Surveys \& Tutorials}}
  \bibinfo{volume}{16}, \bibinfo{number}{4} (\bibinfo{year}{2014}),
  \bibinfo{pages}{2037--2064}.
\newblock


\bibitem[\protect\citeauthoryear{Honda, Nishida, Raiciu, Greenhalgh, Handley,
  and Tokuda}{Honda et~al\mbox{.}}{2011}]%
        {honda2011still}
\bibfield{author}{\bibinfo{person}{Michio Honda}, \bibinfo{person}{Yoshifumi
  Nishida}, \bibinfo{person}{Costin Raiciu}, \bibinfo{person}{Adam Greenhalgh},
  \bibinfo{person}{Mark Handley}, {and} \bibinfo{person}{Hideyuki Tokuda}.}
  \bibinfo{year}{2011}\natexlab{}.
\newblock \showarticletitle{Is it still possible to extend TCP?}. In
  \bibinfo{booktitle}{\emph{IMC'11}}. ACM, \bibinfo{pages}{181--194}.
\newblock


\bibitem[\protect\citeauthoryear{Housley and Droms}{Housley and Droms}{2018}]%
        {Housley_tls:2018}
\bibfield{author}{\bibinfo{person}{R. Housley} {and} \bibinfo{person}{R.
  Droms}.} \bibinfo{year}{2018}\natexlab{}.
\newblock \bibinfo{title}{TLS 1.3 Option for Negotiation of Visibility in the
  Datacenter}.  (\bibinfo{date}{March} \bibinfo{year}{2018}).
\newblock
\newblock
\shownote{Internet draft, draft-rhrd-tls-tls13-visibility-01.}


\bibitem[\protect\citeauthoryear{{IETF doh Working Group}}{{IETF doh Working
  Group}}{[n. d.]}]%
        {doh-wg}
\bibfield{author}{\bibinfo{person}{{IETF doh Working Group}}.}
  \bibinfo{year}{[n. d.]}\natexlab{}.
\newblock \bibinfo{title}{DNS Over HTTPS}.
\newblock
\newblock
\urldef\tempurl%
\url{https://datatracker.ietf.org/wg/doh/about/}
\showURL{%
Retrieved 2018 from \tempurl}


\bibitem[\protect\citeauthoryear{{IETF quic Working Group}}{{IETF quic Working
  Group}}{[n. d.]}]%
        {quic-wg}
\bibfield{author}{\bibinfo{person}{{IETF quic Working Group}}.}
  \bibinfo{year}{[n. d.]}\natexlab{}.
\newblock \bibinfo{title}{QUIC}.
\newblock
\newblock
\urldef\tempurl%
\url{https://datatracker.ietf.org/wg/doh/about/}
\showURL{%
Retrieved 2018 from \tempurl}


\bibitem[\protect\citeauthoryear{{IO Visor Project}}{{IO Visor Project}}{[n.
  d.]}]%
        {eBPF}
\bibfield{author}{\bibinfo{person}{{IO Visor Project}}.} \bibinfo{year}{[n.
  d.]}\natexlab{}.
\newblock \bibinfo{title}{eBPF, extended Berkeley Packet Filter}.
\newblock
\newblock
\urldef\tempurl%
\url{https://www.iovisor.org/technology/ebpf}
\showURL{%
Retrieved 2018 from \tempurl}


\bibitem[\protect\citeauthoryear{ITU}{ITU}{[n. d.]}]%
        {asn1}
\bibfield{author}{\bibinfo{person}{ITU}.} \bibinfo{year}{[n. d.]}\natexlab{}.
\newblock \bibinfo{title}{Introduction to ASN.1}.
\newblock
\newblock
\urldef\tempurl%
\url{https://www.itu.int/en/ITU-T/asn1/Pages/introduction.aspx}
\showURL{%
Retrieved 2018 from \tempurl}


\bibitem[\protect\citeauthoryear{Jaiswal, Iannaccone, Diot, Kurose, and
  Towsley}{Jaiswal et~al\mbox{.}}{2007}]%
        {jaiswal2007measurement}
\bibfield{author}{\bibinfo{person}{Sharad Jaiswal}, \bibinfo{person}{Gianluca
  Iannaccone}, \bibinfo{person}{Christophe Diot}, \bibinfo{person}{Jim Kurose},
  {and} \bibinfo{person}{Don Towsley}.} \bibinfo{year}{2007}\natexlab{}.
\newblock \showarticletitle{Measurement and classification of out-of-sequence
  packets in a tier-1 IP backbone}.
\newblock \bibinfo{journal}{\emph{IEEE/ACM Transactions on Networking (ToN)}}
  \bibinfo{volume}{15}, \bibinfo{number}{1} (\bibinfo{year}{2007}),
  \bibinfo{pages}{54--66}.
\newblock


\bibitem[\protect\citeauthoryear{John, Tafvelin, and Olovsson}{John
  et~al\mbox{.}}{2010}]%
        {john2010passive}
\bibfield{author}{\bibinfo{person}{Wolfgang John}, \bibinfo{person}{Sven
  Tafvelin}, {and} \bibinfo{person}{Tomas Olovsson}.}
  \bibinfo{year}{2010}\natexlab{}.
\newblock \showarticletitle{Passive internet measurement: Overview and
  guidelines based on experiences}.
\newblock \bibinfo{journal}{\emph{Computer Communications}}
  \bibinfo{volume}{33}, \bibinfo{number}{5} (\bibinfo{year}{2010}),
  \bibinfo{pages}{533--550}.
\newblock


\bibitem[\protect\citeauthoryear{Kakhki, Jero, Choffnes, Nita-Rotaru, and
  Mislove}{Kakhki et~al\mbox{.}}{2017}]%
        {kakhki2017taking}
\bibfield{author}{\bibinfo{person}{Arash~Molavi Kakhki},
  \bibinfo{person}{Samuel Jero}, \bibinfo{person}{David Choffnes},
  \bibinfo{person}{Cristina Nita-Rotaru}, {and} \bibinfo{person}{Alan
  Mislove}.} \bibinfo{year}{2017}\natexlab{}.
\newblock \showarticletitle{Taking a long look at QUIC: an approach for
  rigorous evaluation of rapidly evolving transport protocols}. In
  \bibinfo{booktitle}{\emph{Proceedings of the 2017 Internet Measurement
  Conference}}. ACM, \bibinfo{pages}{290--303}.
\newblock


\bibitem[\protect\citeauthoryear{kernel}{kernel}{[n. d.]a}]%
        {perf}
\bibfield{author}{\bibinfo{person}{The~Linux kernel}.} \bibinfo{year}{[n.
  d.]}\natexlab{a}.
\newblock \bibinfo{title}{perf: Linux profiling with performance counters}.
\newblock
\newblock
\urldef\tempurl%
\url{https://perf.wiki.kernel.org/index.php/Main_Page}
\showURL{%
Retrieved 2018 from \tempurl}


\bibitem[\protect\citeauthoryear{kernel}{kernel}{[n. d.]b}]%
        {tcpinfo}
\bibfield{author}{\bibinfo{person}{The~Linux kernel}.} \bibinfo{year}{[n.
  d.]}\natexlab{b}.
\newblock \bibinfo{title}{struct tcp\_info definition}.
\newblock
\newblock
\urldef\tempurl%
\url{https://elixir.bootlin.com/linux/latest/source/include/uapi/linux/tcp.h#L168}
\showURL{%
Retrieved 2018 from \tempurl}


\bibitem[\protect\citeauthoryear{Langley et~al\mbox{.}}{Langley
  et~al\mbox{.}}{2017}]%
        {langley2017quic}
\bibfield{author}{\bibinfo{person}{Adam Langley} {et~al\mbox{.}}}
  \bibinfo{year}{2017}\natexlab{}.
\newblock \showarticletitle{The QUIC Transport Protocol: Design and
  Internet-Scale Deployment}. In \bibinfo{booktitle}{\emph{SIGCOMM '17}}.
  \bibinfo{publisher}{ACM}, \bibinfo{address}{New York, NY, USA},
  \bibinfo{pages}{183--196}.
\newblock
\showISBNx{978-1-4503-4653-5}
\urldef\tempurl%
\url{http://doi.acm.org/10.1145/3098822.3098842}
\showURL{%
\tempurl}


\bibitem[\protect\citeauthoryear{{Let's Encrypt Certificate Authority,
  ISRG}}{{Let's Encrypt Certificate Authority, ISRG}}{2018}]%
        {letsencrypt}
\bibfield{author}{\bibinfo{person}{{Let's Encrypt Certificate Authority,
  ISRG}}.} \bibinfo{year}{April 17, 2018}\natexlab{}.
\newblock \bibinfo{title}{Percentage of Web Pages Loaded by Firefox Using
  HTTPS}.
\newblock
\newblock
\urldef\tempurl%
\url{https://letsencrypt.org/stats/}
\showURL{%
Retrieved April 17, 2018 from \tempurl}


\bibitem[\protect\citeauthoryear{Li, Springer, Bebis, and Gunes}{Li
  et~al\mbox{.}}{2013}]%
        {li2013survey}
\bibfield{author}{\bibinfo{person}{Bingdong Li}, \bibinfo{person}{Jeff
  Springer}, \bibinfo{person}{George Bebis}, {and} \bibinfo{person}{Mehmet~Hadi
  Gunes}.} \bibinfo{year}{2013}\natexlab{}.
\newblock \showarticletitle{A survey of network flow applications}.
\newblock \bibinfo{journal}{\emph{Journal of Network and Computer
  Applications}} \bibinfo{volume}{36}, \bibinfo{number}{2}
  (\bibinfo{year}{2013}), \bibinfo{pages}{567--581}.
\newblock


\bibitem[\protect\citeauthoryear{Luo, Chan, and Chang}{Luo
  et~al\mbox{.}}{2009}]%
        {luo2009design}
\bibfield{author}{\bibinfo{person}{Xiapu Luo}, \bibinfo{person}{Edmond~WW
  Chan}, {and} \bibinfo{person}{Rocky~KC Chang}.}
  \bibinfo{year}{2009}\natexlab{}.
\newblock \showarticletitle{Design and Implementation of TCP Data Probes for
  Reliable and Metric-Rich Network Path Monitoring.}. In
  \bibinfo{booktitle}{\emph{USENIX Annual Technical Conference}}.
\newblock


\bibitem[\protect\citeauthoryear{Mellia}{Mellia}{2002}]%
        {Mellia_tstat:2002}
\bibfield{author}{\bibinfo{person}{Marco Mellia}.}
  \bibinfo{year}{2002}\natexlab{}.
\newblock \showarticletitle{{TCP Statistic and Analysis Tool}}.
\newblock \bibinfo{journal}{\emph{IEEE Network}} \bibinfo{volume}{16},
  \bibinfo{number}{5} (\bibinfo{date}{Sep} \bibinfo{year}{2002}).
\newblock


\bibitem[\protect\citeauthoryear{Microsoft}{Microsoft}{[n. d.]a}]%
        {windows-ewt}
\bibfield{author}{\bibinfo{person}{Microsoft}.} \bibinfo{year}{[n.
  d.]}\natexlab{a}.
\newblock \bibinfo{title}{Network Tracing in Windows 7: Architecture}.
\newblock
\newblock
\urldef\tempurl%
\url{https://msdn.microsoft.com/en-us/library/windows/desktop/dd569137(v=vs.85).aspx}
\showURL{%
Retrieved 2018 from \tempurl}


\bibitem[\protect\citeauthoryear{Microsoft}{Microsoft}{[n. d.]b}]%
        {ntttcp}
\bibfield{author}{\bibinfo{person}{Microsoft}.} \bibinfo{year}{[n.
  d.]}\natexlab{b}.
\newblock \bibinfo{title}{NTTTCP-for-Linux}.
\newblock
\newblock
\urldef\tempurl%
\url{https://github.com/Microsoft/ntttcp-for-linux}
\showURL{%
Retrieved 2018 from \tempurl}


\bibitem[\protect\citeauthoryear{Moshref, Yu, Govindan, and Vahdat}{Moshref
  et~al\mbox{.}}{2016}]%
        {trumpet}
\bibfield{author}{\bibinfo{person}{Masoud Moshref}, \bibinfo{person}{Minlan
  Yu}, \bibinfo{person}{Ramesh Govindan}, {and} \bibinfo{person}{Amin Vahdat}.}
  \bibinfo{year}{2016}\natexlab{}.
\newblock \showarticletitle{Trumpet: Timely and Precise Triggers in Data
  Centers}. In \bibinfo{booktitle}{\emph{SIGCOMM '16}}. \bibinfo{address}{New
  York, NY, USA}, \bibinfo{pages}{129--143}.
\newblock
\showISBNx{978-1-4503-4193-6}


\bibitem[\protect\citeauthoryear{Nguyen and Roughan}{Nguyen and
  Roughan}{2013}]%
        {internet-loss-rates}
\bibfield{author}{\bibinfo{person}{Hung~X Nguyen} {and}
  \bibinfo{person}{Matthew Roughan}.} \bibinfo{year}{2013}\natexlab{}.
\newblock \showarticletitle{Rigorous statistical analysis of internet loss
  measurements}.
\newblock \bibinfo{journal}{\emph{IEEE/ACM Transactions on Networking (TON)}}
  \bibinfo{volume}{21}, \bibinfo{number}{3} (\bibinfo{year}{2013}),
  \bibinfo{pages}{734--745}.
\newblock


\bibitem[\protect\citeauthoryear{Paasch, Barr\'{}e, et~al\mbox{.}}{Paasch
  et~al\mbox{.}}{[n. d.]}]%
        {multipath-tcp-org}
\bibfield{author}{\bibinfo{person}{C. Paasch}, \bibinfo{person}{S. Barr\'{}e},
  {et~al\mbox{.}}} \bibinfo{year}{[n. d.]}\natexlab{}.
\newblock \bibinfo{title}{Multipath TCP in the Linux Kernel}.
\newblock
\newblock
\urldef\tempurl%
\url{https://www.multipath-tcp.org}
\showURL{%
Retrieved 2018 from \tempurl}


\bibitem[\protect\citeauthoryear{Papastergiou, Fairhurst, Ros, Brunstrom,
  Grinnemo, Hurtig, Khademi, T{\"u}xen, Welzl, Damjanovic,
  et~al\mbox{.}}{Papastergiou et~al\mbox{.}}{2017}]%
        {papastergiou2017ossifying}
\bibfield{author}{\bibinfo{person}{Giorgos Papastergiou},
  \bibinfo{person}{Gorry Fairhurst}, \bibinfo{person}{David Ros},
  \bibinfo{person}{Anna Brunstrom}, \bibinfo{person}{Karl-Johan Grinnemo},
  \bibinfo{person}{Per Hurtig}, \bibinfo{person}{Naeem Khademi},
  \bibinfo{person}{Michael T{\"u}xen}, \bibinfo{person}{Michael Welzl},
  \bibinfo{person}{Dragana Damjanovic}, {et~al\mbox{.}}}
  \bibinfo{year}{2017}\natexlab{}.
\newblock \showarticletitle{De-ossifying the internet transport layer: A survey
  and future perspectives}.
\newblock \bibinfo{journal}{\emph{IEEE Communications Surveys \& Tutorials}}
  \bibinfo{volume}{19}, \bibinfo{number}{1} (\bibinfo{year}{2017}),
  \bibinfo{pages}{619--639}.
\newblock


\bibitem[\protect\citeauthoryear{Pearce}{Pearce}{2014}]%
        {blackhat-mptcp}
\bibfield{author}{\bibinfo{person}{K. Pearce}.}
  \bibinfo{year}{2014}\natexlab{}.
\newblock \bibinfo{title}{Multipath TCP Breaking today's networks with
  tomorrow's protocol}.  (\bibinfo{year}{2014}).
\newblock
\newblock
\shownote{Presentation at BlackHat 2014.}


\bibitem[\protect\citeauthoryear{Postel}{Postel}{1981}]%
        {rfc793}
\bibfield{author}{\bibinfo{person}{Jon Postel}.}
  \bibinfo{year}{1981}\natexlab{}.
\newblock \bibinfo{booktitle}{\emph{Transmission Control Protocol}}.
\newblock \bibinfo{type}{STD}~7.
\newblock
\showISSN{2070-1721}


\bibitem[\protect\citeauthoryear{project}{project}{[n. d.]}]%
        {chrome-logger}
\bibfield{author}{\bibinfo{person}{The~Chromium project}.} \bibinfo{year}{[n.
  d.]}\natexlab{}.
\newblock \bibinfo{title}{How to enable loggin in Chromium}.
\newblock
\newblock
\urldef\tempurl%
\url{https://www.chromium.org/for-testers/enable-logging}
\showURL{%
Retrieved 2018 from \tempurl}


\bibitem[\protect\citeauthoryear{Raghunarayan}{Raghunarayan}{2005}]%
        {tcp-mib}
\bibfield{author}{\bibinfo{person}{R. Raghunarayan}.}
  \bibinfo{year}{2005}\natexlab{}.
\newblock \bibinfo{booktitle}{\emph{Management Information Base for the
  Transmission Control Protocol (TCP)}}.
\newblock \bibinfo{type}{RFC} 4022.
\newblock
\showISSN{2070-1721}


\bibitem[\protect\citeauthoryear{Rescorla}{Rescorla}{2018}]%
        {tls13}
\bibfield{author}{\bibinfo{person}{Eric Rescorla}.}
  \bibinfo{year}{2018}\natexlab{}.
\newblock \bibinfo{booktitle}{\emph{The Transport Layer Security (TLS) Protocol
  Version 1.3}}.
\newblock \bibinfo{type}{Internet-Draft} draft-ietf-tls-tls13-28.
  \bibinfo{institution}{IETF}.
\newblock


\bibitem[\protect\citeauthoryear{R{\"u}th, Poese, Dietzel, and
  Hohlfeld}{R{\"u}th et~al\mbox{.}}{2018}]%
        {ruth2018first}
\bibfield{author}{\bibinfo{person}{Jan R{\"u}th}, \bibinfo{person}{Ingmar
  Poese}, \bibinfo{person}{Christoph Dietzel}, {and} \bibinfo{person}{Oliver
  Hohlfeld}.} \bibinfo{year}{2018}\natexlab{}.
\newblock \showarticletitle{A First Look at QUIC in the Wild}. In
  \bibinfo{booktitle}{\emph{International Conference on Passive and Active
  Network Measurement}}. Springer, \bibinfo{pages}{255--268}.
\newblock


\bibitem[\protect\citeauthoryear{Santos}{Santos}{2015}]%
        {santos2015network}
\bibfield{author}{\bibinfo{person}{Omar Santos}.}
  \bibinfo{year}{2015}\natexlab{}.
\newblock \bibinfo{booktitle}{\emph{Network Security with NetFlow and IPFIX:
  Big Data Analytics for Information Security}}.
\newblock \bibinfo{publisher}{Cisco Press}.
\newblock


\bibitem[\protect\citeauthoryear{Schills}{Schills}{[n. d.]}]%
        {packetdrill-mptcp}
\bibfield{author}{\bibinfo{person}{Arnaud Schills}.} \bibinfo{year}{[n.
  d.]}\natexlab{}.
\newblock \bibinfo{title}{Packetdrill test suite for \MPTCP{}}.
\newblock
\newblock
\urldef\tempurl%
\url{https://github.com/aschils/packetdrill_mptcp}
\showURL{%
Retrieved 2018 from \tempurl}


\bibitem[\protect\citeauthoryear{Sekar, Duffield, Spatscheck, van~der Merwe,
  and Zhang}{Sekar et~al\mbox{.}}{2006}]%
        {sekar2006lads}
\bibfield{author}{\bibinfo{person}{Vyas Sekar}, \bibinfo{person}{Nick~G
  Duffield}, \bibinfo{person}{Oliver Spatscheck}, \bibinfo{person}{Jacobus~E
  van~der Merwe}, {and} \bibinfo{person}{Hui Zhang}.}
  \bibinfo{year}{2006}\natexlab{}.
\newblock \showarticletitle{LADS: Large-scale Automated DDoS Detection
  System.}. In \bibinfo{booktitle}{\emph{USENIX Annual Technical Conference,
  General Track}}. \bibinfo{pages}{171--184}.
\newblock


\bibitem[\protect\citeauthoryear{Stephan et~al\mbox{.}}{Stephan
  et~al\mbox{.}}{2017}]%
        {Stephan_QUIC:2017}
\bibfield{author}{\bibinfo{person}{Emile Stephan} {et~al\mbox{.}}}
  \bibinfo{year}{2017}\natexlab{}.
\newblock \bibinfo{title}{{QUIC Interdomain Troubleshooting}}.
  (\bibinfo{date}{July} \bibinfo{year}{2017}).
\newblock
\newblock
\shownote{Internet draft,
  draft-stephan-quic-interdomain-troubleshooting-00.txt, work in progress.}


\bibitem[\protect\citeauthoryear{{The Android project}}{{The Android
  project}}{2018}]%
        {cronet}
\bibfield{author}{\bibinfo{person}{{The Android project}}.}
  \bibinfo{year}{2018}\natexlab{}.
\newblock \bibinfo{title}{Perform network operations using Cronet}.
\newblock
\newblock
\urldef\tempurl%
\url{https://developer.android.com/guide/topics/connectivity/cronet/}
\showURL{%
Retrieved 2018 from \tempurl}


\bibitem[\protect\citeauthoryear{{The IOvisor project}}{{The IOvisor
  project}}{[n. d.]}]%
        {bcc}
\bibfield{author}{\bibinfo{person}{{The IOvisor project}}.} \bibinfo{year}{[n.
  d.]}\natexlab{}.
\newblock \bibinfo{title}{BPF Compiler Collection (bcc)}.
\newblock
\newblock
\urldef\tempurl%
\url{https://www.iovisor.org/technology/bcc}
\showURL{%
Retrieved 2018 from \tempurl}


\bibitem[\protect\citeauthoryear{{The Linux Foundation}}{{The Linux
  Foundation}}{[n. d.]}]%
        {tcpprobe}
\bibfield{author}{\bibinfo{person}{{The Linux Foundation}}.} \bibinfo{year}{[n.
  d.]}\natexlab{}.
\newblock \bibinfo{title}{TCP Probe}.
\newblock
\newblock
\urldef\tempurl%
\url{https://wiki.linuxfoundation.org/networking/tcpprobe}
\showURL{%
Retrieved 2018 from \tempurl}


\bibitem[\protect\citeauthoryear{Trammel et~al\mbox{.}}{Trammel
  et~al\mbox{.}}{2018}]%
        {Trammel_spin:2018}
\bibfield{author}{\bibinfo{person}{B. Trammel} {et~al\mbox{.}}}
  \bibinfo{year}{2018}\natexlab{}.
\newblock \bibinfo{title}{{Adding Explicit Passive Measurability of Two-Way
  Latency to the QUIC Transport Protocol}}.  (\bibinfo{date}{May}
  \bibinfo{year}{2018}).
\newblock
\newblock
\shownote{Internet draft, draft-trammell-quic-spin-03, work in progress.}


\bibitem[\protect\citeauthoryear{Trammell and Boschi}{Trammell and
  Boschi}{2011}]%
        {trammell2011introduction}
\bibfield{author}{\bibinfo{person}{Brian Trammell} {and} \bibinfo{person}{Elisa
  Boschi}.} \bibinfo{year}{2011}\natexlab{}.
\newblock \showarticletitle{An introduction to IP flow information export
  (IPFIX)}.
\newblock \bibinfo{journal}{\emph{IEEE Communications Magazine}}
  \bibinfo{volume}{49}, \bibinfo{number}{4} (\bibinfo{year}{2011}).
\newblock


\bibitem[\protect\citeauthoryear{Trevisan, Finamore, Mellia, Munafo, and
  Rossi}{Trevisan et~al\mbox{.}}{2017}]%
        {Trevisan_Traffic:2017}
\bibfield{author}{\bibinfo{person}{Martino Trevisan},
  \bibinfo{person}{Alessandro Finamore}, \bibinfo{person}{Marco Mellia},
  \bibinfo{person}{Maurizio Munafo}, {and} \bibinfo{person}{Dario Rossi}.}
  \bibinfo{year}{2017}\natexlab{}.
\newblock \showarticletitle{Traffic Analysis with Off-the-Shelf Hardware:
  Challenges and Lessons Learne}.
\newblock \bibinfo{journal}{\emph{IEEE Communications Magazine}}
  (\bibinfo{date}{March} \bibinfo{year}{2017}).
\newblock


\bibitem[\protect\citeauthoryear{Vaere}{Vaere}{2018}]%
        {DeVaere_QUIC:2018}
\bibfield{author}{\bibinfo{person}{Piet~De Vaere}.}
  \bibinfo{year}{2018}\natexlab{}.
\newblock \emph{\bibinfo{title}{{Adding Passive Measurability to QUIC}}}.
\newblock \bibinfo{thesistype}{Master's\ thesis}. \bibinfo{school}{ETH Zurich}.
\newblock
\newblock
\shownote{MA-2017-16.}


\bibitem[\protect\citeauthoryear{Vallina-Rodriguez, Sundaresan, Kreibich, and
  Paxson}{Vallina-Rodriguez et~al\mbox{.}}{2015}]%
        {vallina2015header}
\bibfield{author}{\bibinfo{person}{Narseo Vallina-Rodriguez},
  \bibinfo{person}{Srikanth Sundaresan}, \bibinfo{person}{Christian Kreibich},
  {and} \bibinfo{person}{Vern Paxson}.} \bibinfo{year}{2015}\natexlab{}.
\newblock \showarticletitle{Header enrichment or isp enrichment?: Emerging
  privacy threats in mobile networks}. In \bibinfo{booktitle}{\emph{Proceedings
  of the 2015 ACM SIGCOMM Workshop on Hot Topics in Middleboxes and Network
  Function Virtualization}}. ACM, \bibinfo{pages}{25--30}.
\newblock


\bibitem[\protect\citeauthoryear{van~der Steeg, Hofstede, Sperotto, and
  Pras}{van~der Steeg et~al\mbox{.}}{2015}]%
        {van2015real}
\bibfield{author}{\bibinfo{person}{Dani{\"e}l van~der Steeg},
  \bibinfo{person}{Rick Hofstede}, \bibinfo{person}{Anna Sperotto}, {and}
  \bibinfo{person}{Aiko Pras}.} \bibinfo{year}{2015}\natexlab{}.
\newblock \showarticletitle{Real-time DDoS attack detection for Cisco IOS using
  NetFlow}. In \bibinfo{booktitle}{\emph{Integrated Network Management (IM),
  2015 IFIP/IEEE International Symposium on}}. IEEE, \bibinfo{pages}{972--977}.
\newblock


\bibitem[\protect\citeauthoryear{Viernickel, Froemmgen, Rizk, Koldehofe, and
  Steinmetz}{Viernickel et~al\mbox{.}}{2018}]%
        {viernickel2018multipath}
\bibfield{author}{\bibinfo{person}{Tobias Viernickel},
  \bibinfo{person}{Alexander Froemmgen}, \bibinfo{person}{Amr Rizk},
  \bibinfo{person}{Boris Koldehofe}, {and} \bibinfo{person}{Ralf Steinmetz}.}
  \bibinfo{year}{2018}\natexlab{}.
\newblock \showarticletitle{Multipath QUIC: A Deployable Multipath Transport
  Protocol}. In \bibinfo{booktitle}{\emph{2018 IEEE International Conference on
  Communications (ICC)}}. IEEE, \bibinfo{pages}{1--7}.
\newblock


\bibitem[\protect\citeauthoryear{Weaverworks}{Weaverworks}{[n. d.]}]%
        {tcp-tracer}
\bibfield{author}{\bibinfo{person}{Weaverworks}.} \bibinfo{year}{[n.
  d.]}\natexlab{}.
\newblock \bibinfo{title}{tcptracer-bpf}.
\newblock
\newblock
\urldef\tempurl%
\url{https://github.com/weaveworks/tcptracer-bpf}
\showURL{%
Retrieved 2018 from \tempurl}


\bibitem[\protect\citeauthoryear{Wing and Yourtchenko}{Wing and
  Yourtchenko}{2012}]%
        {happy-eyeballs}
\bibfield{author}{\bibinfo{person}{D. Wing} {and} \bibinfo{person}{A.
  Yourtchenko}.} \bibinfo{year}{2012}\natexlab{}.
\newblock \bibinfo{booktitle}{\emph{Happy Eyeballs: Success with Dual-Stack
  Hosts}}.
\newblock \bibinfo{type}{RFC} 6555. \bibinfo{institution}{RFC Editor}.
\newblock
\showISSN{2070-1721}


\bibitem[\protect\citeauthoryear{Yeganeh, Rejaie, and Willinger}{Yeganeh
  et~al\mbox{.}}{2017}]%
        {yeganeh2017view}
\bibfield{author}{\bibinfo{person}{Bahador Yeganeh}, \bibinfo{person}{Reza
  Rejaie}, {and} \bibinfo{person}{Walter Willinger}.}
  \bibinfo{year}{2017}\natexlab{}.
\newblock \showarticletitle{A view from the edge: A stub-AS perspective of
  traffic localization and its implications}. In
  \bibinfo{booktitle}{\emph{Network Traffic Measurement and Analysis Conference
  (TMA), 2017}}. IEEE, \bibinfo{pages}{1--9}.
\newblock


\bibitem[\protect\citeauthoryear{Yu, Greenberg, Maltz, Rexford, Yuan, Kandula,
  and Kim}{Yu et~al\mbox{.}}{2011}]%
        {snap11}
\bibfield{author}{\bibinfo{person}{Minlan Yu}, \bibinfo{person}{Albert~G
  Greenberg}, \bibinfo{person}{David~A Maltz}, \bibinfo{person}{Jennifer
  Rexford}, \bibinfo{person}{Lihua Yuan}, \bibinfo{person}{Srikanth Kandula},
  {and} \bibinfo{person}{Changhoon Kim}.} \bibinfo{year}{2011}\natexlab{}.
\newblock \showarticletitle{Profiling Network Performance for Multi-tier Data
  Center Applications.}. In \bibinfo{booktitle}{\emph{NSDI}},
  Vol.~\bibinfo{volume}{11}. \bibinfo{pages}{5--5}.
\newblock


\end{thebibliography}
}

\end{document}